\documentclass[a4paper,11pt]{article}
\usepackage{aaskaiid}
\usepackage{orcidlink}
\usepackage{xcolor}
\usepackage{enumitem}
\usepackage{ulem} 
\usepackage{bm}
\usepackage{graphicx}
\usepackage{subcaption}

\title{Cosmology with Intensity Mapping via Statistics Beyond the Power Spectrum in the SKAO Era}

\ShortTitle{21-cm Statistics Beyond Power Spectrum}

\author[1]{Suman Majumdar  \orcidlink{0000-0001-5948-6920}}
\ShortName{Majumdar et al.} 
\author[2]{Debanjan Sarkar \orcidlink{0000-0001-5763-2541}}
\author[1]{Leon Noble \orcidlink{0009-0004-3138-1130}}
\author[1]{Manas Mohit Dosibhatla \orcidlink{0000-0003-4229-2972}}
\author[3]{Zhaoting Chen \orcidlink{0000-0002-4965-8239}}
\author[4]{Bernhard Vos-Ginés \orcidlink{0000-0002-1803-1169}}
\author[1]{Vishrut Pandya \orcidlink{0009-0009-6595-014X}}
\author[5]{Mohd Kamran \orcidlink{0000-0002-0107-9844}}
\author[6]{Sourav Pal \orcidlink{0009-0002-7022-5830}}
\author[7]{Chandra Shekhar Murmu \orcidlink{0000-0002-1818-5440}}
\author[4]{Cora Uhlemann \orcidlink{0000-0001-7831-1579}}
\author[8,34]{Pauline Gorbatchev \orcidlink{0009-0003-9455-2585}}
\author[9]{Jean-Luc Starck \orcidlink{0000-0003-2177-7794}}
\author[10]{José Luis Bernal \orcidlink{0000-0002-0961-4653}}
\author[11,12]{Marta Spinelli \orcidlink{0000-0003-0148-3254}}
\author[13]{Caroline S. Heneka \orcidlink{0000-0001-8883-0583}}
\author[14]{Matteo Viel \orcidlink{0000-0002-2642-5707}}
\author[5]{Gabriella De Lucia \orcidlink{0000-0002-6220-9104}} 
\author[5]{Fabio Fontanot \orcidlink{0000-0003-4744-0188}}
\author[15,16,17,12]{Stefano Camera \orcidlink{0000-0003-3399-3574}}
\author[18]{Steve Cunnington \orcidlink{0000-0001-6594-107X}}
\author[13]{Benedetta Spina \orcidlink{0000-0003-1634-1283}}
\author[35,19,12]{Dionysios Karagiannis \orcidlink{0000-0002-4927-0816}}
\author[1]{Samit Kumar Pal \orcidlink{0000-0002-2271-4165}}
\author[1]{Anshuman Tripathi \orcidlink{0000-0002-5091-9950}}
\author[20]{Sarah Libanore \orcidlink{0000-0002-2284-9190}}
\author[21,22]{Abinash Kumar Shaw \orcidlink{0000-0002-6123-4383}}
\author[23]{Rajesh Mondal \orcidlink{0000-0001-7728-3756}}
\author[1]{Yashrajsinh Mahida \orcidlink{0009-0000-1796-797X}}

\author[24]{Yashar Akrami \orcidlink{0000-0002-2407-7956}}
\author[18]{Philip Bull \orcidlink{0000-0001-5668-3101}}
\author[25]{Chris Clarkson \orcidlink{0000-0001-7363-0722}}
\author[1]{Abhirup Datta \orcidlink{0000-0002-5333-1095}}
\author[26,27,12]{José Fonseca \orcidlink{0000-0003-0549-1614}}
\author[28]{Yin-Zhe Ma \orcidlink{0000-0001-8108-0986}}
\author[12]{Roy Maartens \orcidlink{0000-0001-9050-5894}}
\author[29]{Liantsoa F. Randrianjanahary \orcidlink{0000-0002-7600-7386}}
\author[30]{Martin Sahlen \orcidlink{0000-0003-0973-4804}}
\author[31]{Xin Wang \orcidlink{0000-0002-2472-6485}}
\author[32]{Yougang Wang \orcidlink{0000-0003-0631-568X}}
\author[33]{Jochen Weller \orcidlink{0000-0002-8282-2010}}

\affiliation[1]{Department of Astronomy, Astrophysics \& Space Engineering, Indian Institute of Technology Indore, Indore 453552, India}
\affiliation[2]{Department of Physics and Trottier Space Institute, McGill University, QC H3A 2T8, Canada}
\affiliation[3]{Institute for Astronomy, The University of Edinburgh, Royal Observatory, Edinburgh EH9 3HJ, UK}
\affiliation[4]{Fakultät für Physik, Universität Bielefeld, Postfach 100131, 33501
Bielefeld, Germany}
\affiliation[5]{INAF-Osservatorio Astronomico di Trieste, Via G. B. Tiepolo 11,
34143 Trieste, Italy}
\affiliation[6]{Physics and Applied Mathematics Unit, Indian Statistical Institute, 203 B.T. Road, Kolkata 700108, India}
\affiliation[7]{Astrophysics Research Centre of the Open University (ARCO),
The Open University of Israel, 1 University Road, PO Box 808, Ra’anana 4353701, Israel}
\affiliation[8]{Institutes of Computer Science and Astrophysics, Foundation for Research and Technology - Hellas (FORTH), Greece}
\affiliation[9]{Université Paris-Saclay, Université Paris Cité, CEA, CNRS, AIM, 91191, Gif-sur-Yvette, France}
\affiliation[10]{Instituto de Física de Cantabria (IFCA), CSIC-Univ. de Cantabria, Avda. de los Castros s/n, E-39005 Santander, Spain}
\affiliation[11]{Observatoire de la Côte d’Azur, Laboratoire Lagrange, Bd de l’Observatoire, CS 34229, 06304 Nice cedex 4, France}
\affiliation[12]{Department of Physics \& Astronomy, University of the Western Cape, Cape Town 7535, South Africa}
\affiliation[13]{Institut f\"ur Theoretische Physik, Universit\"at Heidelberg, Philosophenweg 16, 69120 Heidelberg, Germany}
\affiliation[14]{SISSA, International School for Advanced Studies, Via Bonomea 265, 34136 Trieste TS, Italy}
\affiliation[15]{Dipartimento di Fisica, Universit\`a degli Studi di Torino, Via P.\ Giuria 1, 10125 Torino, Italy}
\affiliation[16]{INFN -- Istituto Nazionale di Fisica Nucleare, Sezione di Torino, Via P.\ Giuria 1, 10125 Torino, Italy}
\affiliation[17]{INAF -- Istituto Nazionale di Astrofisica, Osservatorio Astrofisico di Torino, Strada Osservatorio 20, 10025 Pino Torinese, Italy}
\affiliation[18]{Jodrell Bank Centre for Astrophysics, Department of Physics \& Astronomy, The University of Manchester, Manchester M13 9PL, UK}
\affiliation[19]{Dipartimento di Fisica e Scienze della Terra, Università degli Studi di Ferrara, Via Giuseppe Saragat 1, 44122 Ferrara, Italy}
\affiliation[20]{Department of Physics, Ben-Gurion University of the Negev, Be’er Sheva 84105, Israel}
\affiliation[21]{Department of Computer Science, University of Nevada Las Vegas, 4505 S. Maryland Pkwy., Las Vegas, NV 89154, USA}
\affiliation[22]{Max-Planck-Institut für Astrophysik, Garching D-85748, Germany}
\affiliation[23]{Department of Physics, National Institute of Technology
Calicut, Calicut, 673601, Kerala, India}
\affiliation[24]{Instituto de Física Teórica UAM-CSIC, Campus de Cantoblanco, 28049 Madrid, Spain}
\affiliation[25]{Department of Physics \& Astronomy, Queen Mary University of London, London E1 4NS, U.K.}
\affiliation[26]{Instituto de Astrof\'isica e Ci\^encias do Espa\c{c}o, Universidade do Porto CAUP, 4150-762 Porto, Portugal}
\affiliation[27]{Departamento de F\'isica e Astronomia, Faculdade de Ci\^{e}ncias, Universidade do Porto, Rua do Campo Alegre 687, 4169-007 Porto, Portugal}
\affiliation[28]{Department of Physics, Stellenbosch University, Matieland 7602, South Africa}
\affiliation[29]{Astrophysics Research Centre \& School of Mathematics, Statistics and Computer Science, University of KwaZulu-Natal, Durban, 4041, South Africa}
\affiliation[30]{Department of Physics and Astronomy, Uppsala University, Box 516, 751 20 Uppsala, Sweden}
\affiliation[31]{School of Physics and Astronomy, Sun Yat-Sen University, No. 2 Daxue Rd., Zhuhai 519082, China}
\affiliation[32]{National Astronomical Observatories, Chinese Academy of Sciences, Beijing 100012, China}
\affiliation[33]{Universitäts-Sternwarte München, Fakultät für Physik, Ludwig-Maximilians-Universität München, Scheinerstrasse 1, 81679 München, Germany}
\affiliation[34]{Department of Physics, University of Crete, Greece}
\affiliation[35]{Van Swinderen Institute, University of Groningen, Nijenborgh 3, 9747 AG Groningen, The Netherlands}

\emailAdd{mid.suman@gmail.com}
\emailAdd{debanjan.sarkar@mcgill.ca}

\abstract{
The cosmological distribution of neutral hydrogen (HI) during the post-reionization era is highly non-Gaussian due to the underlying non-linear structure formation, complex galaxy biasing, and potential primordial non-Gaussianity. One needs higher-order (beyond two-point) statistics to maximally extract the non-Gaussian information out of the 21-cm intensity maps. This chapter summarizes the potential of several higher-order statistics, including voxel intensity distribution, emission line stacking, probability density functions, $\ell_1$-norm, bispectrum, and various marked statistics. 
Additionally, image-based morphological descriptors, such as the Largest Cluster Statistic, local dimensions, and Minkowski functionals, etc., can potentially characterize the morphology and geometry of the cosmic web encoded in the 21-cm intensity maps. This chapter presents forecasts of the detectability of these higher-order statistics in the context of the future SKAO observations. These forecasts incorporate instrumental noise, observational effects, and, in some cases, foreground removal in their analyses. With its unprecedented sensitivity, the future SKAO 21-cm observations will enable us to measure these higher-order statistics more precisely, possibly helping to break degeneracies between astrophysical and cosmological parameters, and maximizing the science outcome from these surveys.}

\begin{document}
\newcommand{\actaa}{Acta Astron.} 
\newcommand{\araa}{Annu. Rev. Astron. Astrophys.} 
\newcommand{\aar}{Astron. Astrophys. Rev.} 
\newcommand{\ab}{Astrobiol.} 
\newcommand{\aj}{Astron. J.} 
\newcommand{\apj}{Astrophys. J.} 
\newcommand{\apjl}{Astrophys. J. Lett.} 
\newcommand{\apjs}{Astrophys. J. Suppl. Ser.} 
\newcommand{\ao}{Appl. Opt.} 
\newcommand{\apss}{Astrophys. Space Sci.} 
\newcommand{\aap}{Astron. Astrophys.} 
\newcommand{\aapr}{Astron. Astrophys. Rev.} 
\newcommand{\aaps}{Astron. Astrophys. Suppl.} 
\newcommand{\baas}{Bull. Am. Astron. Soc.} 
\newcommand{\caa}{Chinese Astron. Astrophys.} 
\newcommand{\cjaa}{Chinese J. Astron. Astrophys.} 
\newcommand{\cqg}{Class. Quantum Gravity} 
\newcommand{\gal}{Galaxies} 
\newcommand{\gca}{Geochim. Cosmochim. Acta} 
\newcommand{\icarus}{Icarus} 
\newcommand{\jcap}{J. Cosmol. Astropart. Phys.} 
\newcommand{\jgr}{J. Geophys. Res.} 
\newcommand{\jgrp}{J. Geophys. Res.: Planets} 
\newcommand{\jqsrt}{J. Quant. Spectrosc. Radiat. Transf.} 
\newcommand{\memsai}{Mem. Soc. Astron. Italiana} 
\newcommand{\mnras}{Mon. Not. R. Astron. Soc.} 
\newcommand{\nat}{Nature} 
\newcommand{\nastro}{Nat. Astron.} 
\newcommand{\ncomms}{Nat. Commun.} 
\newcommand{\nphys}{Nat. Phys.} 
\newcommand{\na}{New Astron.} 
\newcommand{\nar}{New Astron. Rev.} 
\newcommand{\physrep}{Phys. Rep.} 
\newcommand{\pra}{Phys. Rev. A} 
\newcommand{\prb}{Phys. Rev. B} 
\newcommand{\prc}{Phys. Rev. C} 
\newcommand{\prd}{Phys. Rev. D} 
\newcommand{\pre}{Phys. Rev. E} 
\newcommand{\prl}{Phys. Rev. Lett.} 
\newcommand{\psj}{Planet. Sci. J.} 
\newcommand{\planss}{Planet. Space Sci.} 
\newcommand{\pnas}{Proc. Natl Acad. Sci. USA} 
\newcommand{\procspie}{Proc. SPIE} 
\newcommand{\pasa}{Publ. Astron. Soc. Aust.} 
\newcommand{\pasj}{Publ. Astron. Soc. Jpn} 
\newcommand{\pasp}{Publ. Astron. Soc. Pac.} 
\newcommand{\rmxaa}{Rev. Mexicana Astron. Astrofis.} 
\newcommand{\sci}{Science} 
\newcommand{\sciadv}{Sci. Adv.} 
\newcommand{\solphys}{Sol. Phys.} 
\newcommand{\sovast}{Soviet Ast.} 
\newcommand{\ssr}{Space Sci. Rev.} 
\newcommand{\uni}{Universe} 

\maketitle
\tableofcontents

\section{Introduction}



The past few decades have established intensity mapping~\citep{Bharadwaj_2001, Bharadwaj_sethi_2001, 2017Kovetz, 2022Bernal} as a transformative observational technique for probing the large-scale structure of the Universe across cosmic time. Line Intensity Mapping (LIM) relies on mapping the integrated line emission from unresolved galaxies and diffuse gas on a coarse resolution element. LIM thus circumvents the observational challenges of individual source detection, granting access to populations of faint and unresolved galaxies and enabling wide-area, deep surveys that traditional galaxy surveys cannot achieve. The 21-cm hyperfine transition from neutral hydrogen (HI)~\citep{Pritchard_2012}, in particular, offers an exceptional tracer of the cosmic matter distribution during the post-reionization epoch ($z \lesssim 6$), with sufficient brightness and abundance to enable high-fidelity mapping across significant cosmic volumes.

The Square Kilometre Array Observatory (SKAO) brings a watershed moment for 21-cm intensity mapping. With its unprecedented sensitivity, collecting area, and redshift coverage spanning $z \approx 0.2$ to $z \approx 3$ with the SKA-Mid array, the SKAO will deliver 21-cm intensity maps at resolutions and signal-to-noise levels far exceeding the capabilities of presently operating radio interferometers. This observational leap creates both opportunity and necessity: while the SKAO's raw sensitivity is formidable, maximizing the scientific return from these observations requires moving beyond traditional two-point statistics to extract the full information content encoded in the data.

Traditionally, LIM analyses have relied primarily on two-point statistics, most commonly the power spectrum, to extract information about the underlying matter field and astrophysical processes. However, the power spectrum is a complete statistical descriptor only in the case of Gaussian random fields. In reality, the late-time matter and HI density fields are highly non-Gaussian due to non-linear structure formation, complex galaxy biasing, astrophysics, and, potentially, primordial non-Gaussianities (PNG) sourced during inflation. The Gaussian assumption, therefore, limits the information extracted from the data.



Higher-order statistics (HOS) of 21-cm LIM are thus essential to overcome the limitations of inference via power spectrum alone and to fully exploit the non-Gaussian information encoded in LIM observations. These HOS include statistics defined in both real and Fourier space. For instance, the bispectrum~\citep{Sarkar_2019, Cunnington_2021, Karagiannis:2022ylq} (three-point correlation function in Fourier space) captures information about mode coupling and the shape dependence of non-Gaussian features, while one-point statistics such as the voxel intensity distribution (VID)~\citep{Breysse:2016szq, Bernal:2023ovz} encode the full temperature PDF of intensity maps, sensitive to the underlying halo mass function and line-luminosity relation. Similarly, marked statistics \citep{Kamran_2025, massara_cosmological_2023} and local morphological descriptors--- such as Largest Cluster Statistics~(LCS; {\citealt{klypin_1993_percolation, BagMondal_2018}}), Minkowski functionals~\citep{mecke_robust_1993, BagMondal_2018, Pathak_2022}, local dimension \citep{sarkar_2009_locdim, dosibhatla_2025_lss-morphology}, and tidal tensor \citep{hahn_properties_2007} --- probe the geometry and topology of the signal beyond amplitude correlations.


The additional information encoded in HOS helps in tightening constraints on cosmological parameters and breaking degeneracies between them. There are also degeneracies between astrophysical and cosmological parameters, for instance, the mean brightness temperature of a target line and the bias with which it traces the matter distribution \citep{castorina_2019_mean-bias-degeneracy}. This can be partly broken using the VID \citep{Breysse:2022alx}. The 21-cm bispectrum has been demonstrated to break the degeneracies of the Hubble parameter ($h$) with evolving dark energy and also primordial power spectrum parameters \citep{Randrianjanahary:2023rgp, pinheiro_2026_degeneracy}. The bispectrum is also an effective summary statistic for breaking the degeneracies of the neutrino mass $M_\nu$ with other cosmological parameters, especially $\sigma_8$ and $\Omega_M$ \citep{hahn_2020_degeneracy}.
Adding the information of the halo mass function and the void size function to the power spectrum can break multiple cosmological degeneracies, including those of $\sigma_8$ with $\Omega_M$, $\Omega_b$, $h$, and $n_s$ \citep{bayer_2021_degeneracies}.
Similarly, segmenting the cosmic web into its constituent nodes, filaments, walls, and voids with the tidal tensor and using their individual power spectra jointly has the potential of breaking the stiff degeneracy between $M_\nu$ and $\sigma_8$ \citep{sunseri_2025_cosmic-web}.

In this chapter, we provide a structured overview of the most promising higher-order statistical tools applicable to 21-cm LIM in the post-reionization regime, emphasizing their complementarity with the power spectrum and their capacity to break degeneracies in cosmological inference. Recent advances in simulation-based inference (SBI) approaches, which circumvent explicit definition of the likelihood function, are crucial for inference using HOS, as these statistics are not expected to have simple Gaussian likelihoods \citep{sun_2026_LIMFAST_SBI, modi_2025_galaxy_SBI, kanafi_2026_SBI, lemos_2023_SBI}. Section \ref{sec:def} outlines the theoretical formalisms of various statistics, including image-domain estimators like the VID, LCS, and local dimension, as well as Fourier-space polyspectra such as the bispectrum and trispectrum. In Section \ref{sec:science}, we present forecasts regarding the detectability of each of these statistics via SKAO alongside their modeling approaches and noise/systematics considerations. For the forecast, we use the SKA-Mid AA4 array layout~\citep{seethapuram_sridhar_2025_16951088} as our reference instrumental configuration. The uncertainty in measuring each statistic from observations can arise from cosmic variance, thermal noise of the radio telescope, residual foregrounds left in the 21-cm maps after foreground cleaning, and various instrumental systematics. Our reference setup considers only the uncertainty due to thermal noise, unless specified otherwise. The required observation time and observation mode for each summary statistic are indicated in the respective sections.

\section{Different Higher-order Statistics of the observed HI 21-cm field}
\label{sec:def}
The observed 21-cm brightness temperature field can be expressed as a smooth background tracing the cosmic mean HI abundance plus a fluctuating component arising from spatial variations in the HI distribution~\citep{Pritchard:2008da, pritchard201221, Furlanetto:2009qk, Furlanetto:2006jb}.  
In the post-reionization universe, the intergalactic medium is highly ionized, and neutral hydrogen survives only in dense, self-shielded regions such as galaxies and damped Ly$\alpha$ systems.
The 21-cm brightness temperature thus can be expressed as
\begin{equation}
T_b(z) = \bar{T}_b(z)\,[1+\delta_{\rm HI}(z)],
\end{equation}
where $\delta_{\rm HI}$ is the HI overdensity and the mean brightness temperature is given by~\citep{battye2012bingo, Bull:2014rha, Camera:2015yqa, Padmanabhan:2023hfr}
\begin{equation}
\bar{T}_b(z) = 0.566\,h\,\frac{H_0}{H(z)}\!\left(\frac{\Omega_{\rm HI}}{0.003}\right)\!(1+z)^2~{\rm mK}.
\end{equation}
Observations and simulations~\citep{Padmanabhan:2023hfr,Castorina:2016bfm,Villaescusa-Navarro:2014cma} indicate that $\Omega_{\rm HI}$ is approximately constant for $0 \lesssim z \lesssim 6$, with $\Omega_{\rm HI} \simeq 4.7\times10^{-4}$.

\subsection{One-point statistics} 

In cosmology, the probability distribution of matter density fluctuations is key to understanding the large-scale structure of the universe. While linear theory predicts Gaussian statistics for early-universe fluctuations, gravitational evolution induces significant non-Gaussianities at late times. 

The voxel intensity distribution (VID) is an estimator of the probability distribution function of the measured temperatures in a given voxel~\citep{Breysse:2016szq, Bernal:2023ovz}. Intuitively, such probability $\mathcal{P}(T)$ is the marginalization over the conditional probability $\mathcal{P}(T|N)$ of measuring a temperature $T$ in a voxel with $N$ galaxies:
\begin{equation}
	\mathcal{P}(T) = \sum_N \mathcal{P}(T|N)\mathcal{P}(N)\,,
	\label{eq:VID}
\end{equation}
where $\mathcal{P}(N)$ is the probability of having $N$ galaxies contributing within a voxel. Therefore, besides granting access to the non-Gaussian information in line-intensity maps, the VID is very sensitive to the line-luminosity function and hence highly complementary to the power spectrum. While any modeling of the VID conceptually reduces to Eq.~\eqref{eq:VID}, there are different approaches~\citep{Bernal:2023ovz} for its computation. If properly combined with the power spectrum~\citep{COMAP:2018kem, Sato-Polito:2022fkd}, the VID breaks degeneracies between astrophysical and cosmological uncertainties (see e.g.,~\citealt{Breysse:2022alx}), boosting the constraining power of LIM for beyond $\Lambda$CDM models~\citep{Sabla:2024dxz}. 

While the VID contains a wealth of information, it is also highly susceptible to observational systematics. There are different approaches involving cross-correlations at one-point that increase the robustness of these measurements. While the deconvolved density estimator~\citep{Breysse:2022fdi, COMAP:2022sdg} and the conditional VID~\citep{Breysse:2019cdw} aim to capture the probability distributions of different tracers, stacking averages the LIM signal around known tracers~\citep{LujanNiemeyer:2022rby, MeerKLASS:2025lmz, Dunne:2025uvc}. Given its robustness and higher signal-to-noise ratio, the latter is optimal to characterize the signal in noisier observations and to study instrumental and observational systematics, and therefore can be a preparatory step for more complex studies of one-point statistics. In this subsection, we discuss the emission line stacking of the 21-cm signals. The stacked signal is calculated as 
\begin{equation}
    I(\Delta\alpha,\Delta\phi,\Delta\nu) = \frac{\sum_i I(\alpha_i{+}\Delta\alpha,\,\phi_i{+}\Delta\phi,\,\nu_i{+}\Delta\nu) w_i}{\sum_i w_i},
\label{eq:stack}
\end{equation} 
where $(\alpha_i,\phi_i)$ are the R.A. and decl. of the voxel in which the $i^{\rm th}$ galaxy resides, $I$ is the flux density, and $w_i$ is the weight of each voxel, $\nu_i$ is the $i^{\rm th}$ frequency channel and $\Delta \nu$ is the channel width.

Finally, Large Deviation Theory (LDT) offers a robust framework to model the PDF of densities in spheres by capturing the exponentially rare but physically relevant fluctuations driven by nonlinear gravitational dynamics \citep{Uhlemann2016}. In this context, the PDF is approximated as:

\begin{equation}
    \mathcal{P}_R(\rho) \propto \exp\left[-\Psi_R(\rho)\right], \quad \Psi_R (\rho)= \frac{1}{2} \frac{\delta_{\rm L}^2}{\sigma_{\rm L}^2(R\rho^{1/3})} \frac{\sigma_{\rm L}^2(R)}{\sigma_{\rm NL}^2(R)} 
\end{equation}

where $\rho = 1+\delta$, $\Psi_R (\rho)$ is the rate function computed at a smoothing radius $R$, as a function of the linear variance $\sigma_{\rm L}^2$, the non-linear variance $\sigma_{\rm NL}^2$ and the linear overdensity $\delta_{\rm L}(\rho)$, which is usually related to the late-time density $\rho$ using spherical collapse theory. This formulation enables accurate predictions of the PDF into the non-Gaussian regime, provided that the non-linear variance is well below unity. The PDF of biased tracers such as neutral hydrogen \citep{Leicht2019} can be modeled considering bias and shot-noise parametrizations in terms of the dark matter overdensity \citep{Gould2025}.

\subsection{Bispectrum}
The post-reionization HI 21-cm signal is expected to be significantly non-Gaussian on length scales that have become non-linear.
The power spectrum, by definition, measures the amplitude of the signal fluctuations at different scales and only completely characterizes the statistical properties of a signal if it is a Gaussian random field. In order to quantify the non-Gaussianity present in the 21-cm signal, one has to choose a statistic sensitive to that non-Gaussianity. One-point statistics, such as skewness and kurtosis, are such possibilities. However, these one-point statistics can only capture the non-Gaussianity 
globally and fail to capture the correlations between the non-Gaussianity present at different scales. To comprehensively quantify both, one has to consider higher-order statistics such as the bispectrum. The bispectrum $B_{\rm 21-cm}(\mathbf{k_1},\mathbf{k_2},\mathbf{k_3})$ of the 21-cm  differential brightness temperature field $\delta T_{b}(\mathbf{x})$ is defined as 
\begin{align}
    \langle \Delta_{b}(\mathbf{k_1}) \Delta_{b}(\mathbf{k_2}) \Delta_{b}(\mathbf{k_3}) \rangle =V \delta_{\mathbf{k_1} + \mathbf{k_2} + \mathbf{k_3},0}~B_{\rm 21-cm}(\mathbf{k_1},\mathbf{k_2},\mathbf{k_3}),
\end{align}
where the $\Delta_{b}(\mathbf{k})$ is the Fourier transform of the differential brightness temperature field $\delta T_{b}(\mathbf{x})$ and V is the total volume of the simulation box or observation volume. The $\delta_{\mathbf{k_1} + \mathbf{k_2} + \mathbf{k_3},0}$ is the Kronecker delta function, and its numerical value is equal to one when the condition  $\mathbf{k_1} + \mathbf{k_2} + \mathbf{k_3}=0$ is satisfied and zero otherwise. This ensures that only the closed $k$-triangles contribute to the bispectrum estimates. The bispectrum can be estimated for different sizes and shapes of the $k-\text{triangles}$ in Fourier space. To identify all of the unique shapes of the triangles in the Fourier space, one can use the bispectrum parametrization introduced in \cite{Bharadwaj_2020}. According to this parametrization, for a triangle in Fourier space with $k_1\leq k_2 \leq k_3$, its size is determined by $k_1$, and the shape is determined by $t=k_2/k_1$ and $\mu$, the cosine of the angle  ($\cos \theta$) between $k_1$ and $k_2$.

\subsubsection{Bispectrum multipole moments}
The anisotropies due to the redshift space distortions (RSD) in the 21-cm bispectrum contain a wealth of cosmological information, which is completely quantified using HI bispectrum multipole moments. As RSD introduces angular dependence in the bispectrum with respect to the line of sight, the redshift-space bispectrum is expanded in spherical harmonics as~\citep{Bharadwaj_2020}
\begin{equation}
B_{\rm HI}^s(k_1,\mu,t,\hat{\mathbf{p}}) = 
\sum_{\ell=0}^\infty \sum_{m=-\ell}^{\ell} 
B_\ell^m(k_1,\mu,t)\,Y_\ell^m(\hat{\mathbf{p}}),
\end{equation}
where $B_\ell^m$ are the bispectrum multipoles and $\hat{\mathbf{p}}$ characterizes the orientation of the triangle relative to the line of sight.  
To highlight anisotropic effects, we define the reduced multipoles
\begin{equation}
Q_\ell^m(k_1,\mu,t,z) \equiv 
\frac{B_\ell^m(k_1,\mu,t,z)}{B_{\rm HI}^r(k_1,\mu,t,z)},
\end{equation}
which remove the overall amplitude $\bar{T}_b^3$ and emphasize the RSD- and Finger-of-God (FoG)-induced anisotropies.

\subsubsection{Modeling the HI bispectrum}
On large scales, the HI field can be modeled as a locally biased tracer of the matter overdensity:
\begin{equation}
\delta_{\rm HI} = b_1\,\delta_{\rm m} + \frac{b_2}{2}\,\delta_{\rm m}^2 ,
\end{equation}
where $b_1$ and $b_2$ are the linear and quadratic bias parameters, respectively.  
We adopt redshift-dependent fits calibrated from simulated real-space HI bispectrum measurements~\citep{Sarkar:2016lvb, Sarkar:2019ojl}.  
At tree level in perturbation theory, the real-space HI bispectrum for a closed Fourier triangle $(\mathbf{k}_1,\mathbf{k}_2,\mathbf{k}_3)$ can be written schematically as
\begin{equation}
B_{\rm HI}(k_1,k_2,k_3) = \frac{1}{b_1}\Big[2F_2\,P_{\rm HI}(k_1)P_{\rm HI}(k_2)+\text{cyc.}\Big]
+ \frac{b_2}{b_1^2}\Big[P_{\rm HI}(k_1)P_{\rm HI}(k_2)+\text{cyc.}\Big],
\end{equation}
where $F_2$ is the standard symmetrized second-order kernel.  
The first term describes non-linear gravitational mode coupling (shape-dependent), while the second captures isotropic contributions from non-linear HI bias. As these two terms have distinct geometric dependences, measurements of the bispectrum can independently constrain $b_1$ and $b_2$, improving upon power-spectrum-only analyses.

\subsection{Marked Power Spectrum}

Traditional two-point statistics, such as the power spectrum, are optimal for Gaussian random fields; however, the late-time large-scale structure of the Universe is highly non-Gaussian due to gravitational evolution. As structures collapse and form filaments, clusters, and voids, higher-order correlations emerge that the standard power spectrum cannot capture. Marked statistics, as introduced in \cite{White_2016}, address this limitation by introducing a non-linear transformation—a `mark'—applied to the density field before computing its two-point function. This mark is typically a function of local environmental properties such as smoothed brightness temperature, density, halo mass, or velocity dispersion, and it effectively re-weights different regions of space to enhance sensitivity to specific physical regimes. This approach provides an independent and computationally cheaper alternative to the higher-order statistics to describe the non-Gaussianity in the field.

The Marked Power spectrum $\mathbf{P_M(k)}$ of a field $A(\mathbf{x})$ can be defined as

\begin{equation} \label{marked_power_spectrum}
    \langle m_{A}(\mathbf{k}) \, m_{A}^*(\mathbf{k'}) \rangle = V \, \delta_\mathbf{k + k'} \, \mathbf{P_M (k)},
\end{equation}

where $V$ is the simulation volume and $m_A(\mathbf{k})$ is the Fourier transform of the mark $m_A{(\mathbf{x})}$-a function of $A(\mathbf{x})$, acting as a transformation. 

Recent works by \citet{Kamran_2025} and \citet{Pandya:2026chj} have demonstrated that applying marked statistics to simulated 21-cm fields can enhance sensitivity to the evolving spatial distribution of HI across redshift. For instance, weighting the HI field by inverse powers of the local smoothed density suppresses contributions from overdense, highly biased regions, thereby amplifying the signal from under-dense environments that are otherwise washed out in unmarked analyses. This reweighting helps to probe the environmental dependence of HI clustering, trace changes in the cosmic HI reservoir, and break degeneracies in cosmological parameter inference. As future 21-cm surveys such as SKA-Mid target the post-reionization era, marked statistics will be an essential tool to access higher-order information encoded in the large-scale distribution of neutral hydrogen.

\subsection{Starlet $\ell_1$-norm}

The $\ell_1$-norm of isotropic undecimated wavelet coefficients offers a compact, multiscale summary of HI intensity maps, sensitive to sparse, non-Gaussian structures not captured by traditional two-point statistics.

Let \( M(\hat{n}) \) denote the HI temperature fluctuation map on the sphere, where \( \hat{n} \) is the sky direction. The spherical Starlet transform decomposes this map into a set of wavelet coefficients \( w_j(\hat{n}) \) at each scale \( j = 1, \dots, J \), and a coarse residual map \( c_J(\hat{n}) \), as follows:

\begin{equation}
M(\hat{n}) = \sum_{j=1}^J w_j(\hat{n}) + c_J(\hat{n}).
\label{eq:Starlet_decomposition}
\end{equation}

Each wavelet scale \( w_j(\hat{n}) \) captures angular features localized in a specific band of multipoles. The coefficients are normalized by a scale-dependent factor \( \mathcal{N}_j \) to account for varying amplitudes across scales. The normalized coefficients are then binned in signal-to-noise ratio (SNR) space.

We define the $\ell_1$-norm of the normalized coefficients at scale \( j \) within SNR bin \( k \) as:

\begin{equation}
\ell_1^{(j,k)} = \sum_{\hat{n} \in \mathcal{B}_k} \left| \frac{w_j(\hat{n})}{\mathcal{N}_j} \right|,
\label{eq:l1_norm}
\end{equation}
where \( \mathcal{B}_k \) is the set of pixels whose normalized coefficient falls in the \( k \)-th bin:

\begin{equation}
\mathcal{B}_k = \left\{ \hat{n} \,\middle|\, T_k \leq \frac{w_j(\hat{n})}{\mathcal{N}_j} < T_{k+1} \right\},
\label{eq:l1_pixels}
\end{equation}
and \( T_k \) are the SNR bin edges. The full summary statistic is \( \ell_1^{(j,k)} \in \mathbb{R}^{J \times K} \), where \( K \) is the number of bins and \( J \) is the number of wavelet scales.

\subsection{Morphological Measures in the Image Domain}
\label{subsec:image_statistics_formalism}

Even though marked statistics and polyspectra retain the non-Gaussian information in intensity maps, they cannot preserve the complete phase information in Fourier space, resulting in information loss of morphological features of the LSS. This motivates analyzing intensity maps in the image domain to characterize their topology and morphology.

\subsubsection{Largest Cluster Statistic}
\label{subsubsec:LCS_formalism}

In a binary field that can be segmented into clusters and regions not belonging to any cluster, the largest cluster statistic (LCS) is defined as the fraction of cluster volume that resides in the largest cluster \citep{klypin_1993_percolation, Bharadwaj_2000, BagMondal_2018, Pathak_2022, Dasgupta_2023, pal_2025_lcs}:
\begin{equation}
    {\rm LCS} = \frac{\text{Volume of the largest cluster}}{\text{Total cluster volume}} \, .
\end{equation}
In line intensity maps targeting emission lines from galaxies, a cluster can be defined as a connected set of bright spots in the line emission.

Percolation transition refers to the onset of large-scale connectivity in the cosmic web, where the matter distribution connects to form one contiguous large cluster spanning the whole universe. The LCS is useful for studying the percolation transition of an intensity map \citep{shandarin_1983_percolation, klypin_1993_percolation, Bharadwaj_2000, Pathak_2022, Dasgupta_2023, Regos_2024, pal_2025_lcs}, which is quantified by the evolution of the LCS with the filling factor (FF) of the map, defined as
\begin{equation}
    {\rm FF} = \frac{\text{Total cluster volume}}{\text{Total survey volume}} \, .
\end{equation}

As the bright spots grow due to the growth of structure, the cluster volume, and hence the filling factor, grow steadily. However, at the stage where large clusters start to merge, the volume of the largest cluster grows sharply and becomes comparable to the total cluster volume. This makes the LCS approach $1$ abruptly and marks the percolation transition.

\begin{figure}[htbp]
    \centering
    \includegraphics[width=\textwidth]{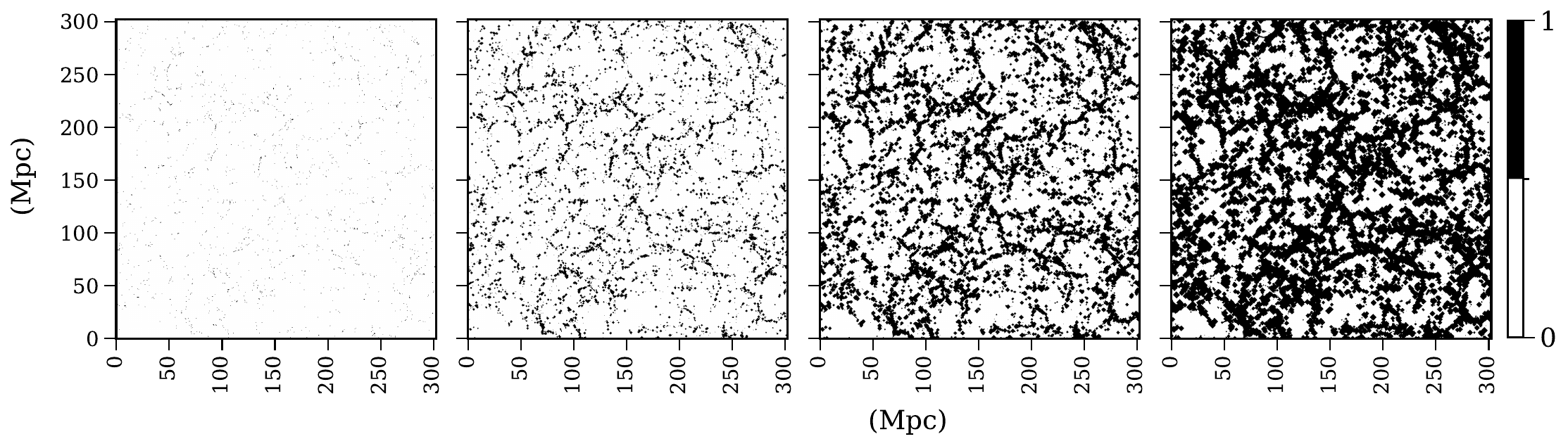}
    \caption{Illustration of the iterative coarse-graining scheme used for percolation analysis. The four panels show the same slice of a field without coarse-graining (first from left) and after 2, 4, and 6 iterations in order. On each iteration, the cells sharing a face with a bright cell are marked as bright. The fields are binary, with the dark cells marked by zeros and the bright cells by ones. See \cite{dosibhatla_2025_lss-morphology} for more details.
    \label{fig:coarse-grain}}
\end{figure}

In practice, the resolution of intensity maps may not be adequate to reveal the connectivity of the matter distribution, as intensity maps trace line emission from biased tracers (galaxies and diffuse gas). Therefore, we employ an iterative coarse-graining scheme as shown in Figure \ref{fig:coarse-grain}, where cells adjacent to bright cells are marked as bright \citep{Bharadwaj_2000, dosibhatla_2025_lss-morphology}. This is done repeatedly, and the FF and LCS are calculated after every iteration. The evolution of the LCS with FF gives the percolation curve. The percolation transition is expected to be attained at different filling factors for different intensity maps, and therefore, the LCS is a useful higher-order summary statistic.

\subsubsection{Local Dimension}
\label{subsubsec:local_dimension_formalism}

The LSS of the universe consists of filaments, sheets, nodes, and voids, which form a cellular structure known as the ``cosmic web'' \citep{gregory_1978_cosmic-web, einasto_1984_cosmic-web, zeldovich_1982_giant-voids, bond_1996_cosmic-web, Bharadwaj_2000, aragon-calvo_2010_cosmic-web, libeskind_2017_cosmic-web, tojeiro_2025_cosmic-web_review}. The contribution of the different building blocks of the cosmic web to the line emission in an intensity map contains a wealth of non-Gaussian and morphological information. The local dimension \citep{sarkar_2009_locdim, sarkar_2012_locdim, sarkar_2019_locdim, pandey_2020_locdim} is a simple yet effective way of determining whether a cell in an intensity map belongs to a filament, sheet, or a volume-filling environment (node or void) \citep{dosibhatla_2025_lss-morphology}. The local dimension ($D$) of a cell in an intensity map determines how the number of bright cells, $N(R)$, in a sphere centered at the given bright cell scales with the radius of the sphere.
\begin{equation}
    N(R) = AR^D
    \label{eq:local_dimension}
\end{equation}
where $A$ is a normalization constant. For straight filaments and plane sheets passing through the center of the sphere, and volume-filling environments, $D$ takes values $\simeq$ $1$, $2$, and $3$ respectively. However, fractional values are allowed for environments intermediate between any two kinds.

To compute the local dimension of any bright cell in an intensity map, we count $N(R)$ by incrementing $R$ by a multiple $\Delta R$ of the grid resolution between fixed values $R_{\rm min}$ and $R_{\rm max}$. We then fit the variation of $N(R)$ with $R$ using equation \ref{eq:local_dimension}. If the fit converges with $\chi^2_\nu \leq \chi^2_{\nu,{\rm max}}$, the cell is classifiable, and it belongs to the environment corresponding to $D$ in the length scale $R_{\rm min}$ to $R_{\rm max}$. The value of $\chi^2_{\nu,{\rm max}}$ is chosen such that a good fit is obtained and a large number of cells is classifiable. The fraction of cells in different environments characterizes the morphology of an intensity map effectively.

\section{Science with Higher-Order Statistics}
\label{sec:science}

\subsection{One-point statistics}
\label{sec:One_point_statistics_science}

\begin{figure}[htbp]
    \centering
    \begin{subfigure}[t]{0.61\textwidth}
        \centering
        \includegraphics[width=\textwidth]{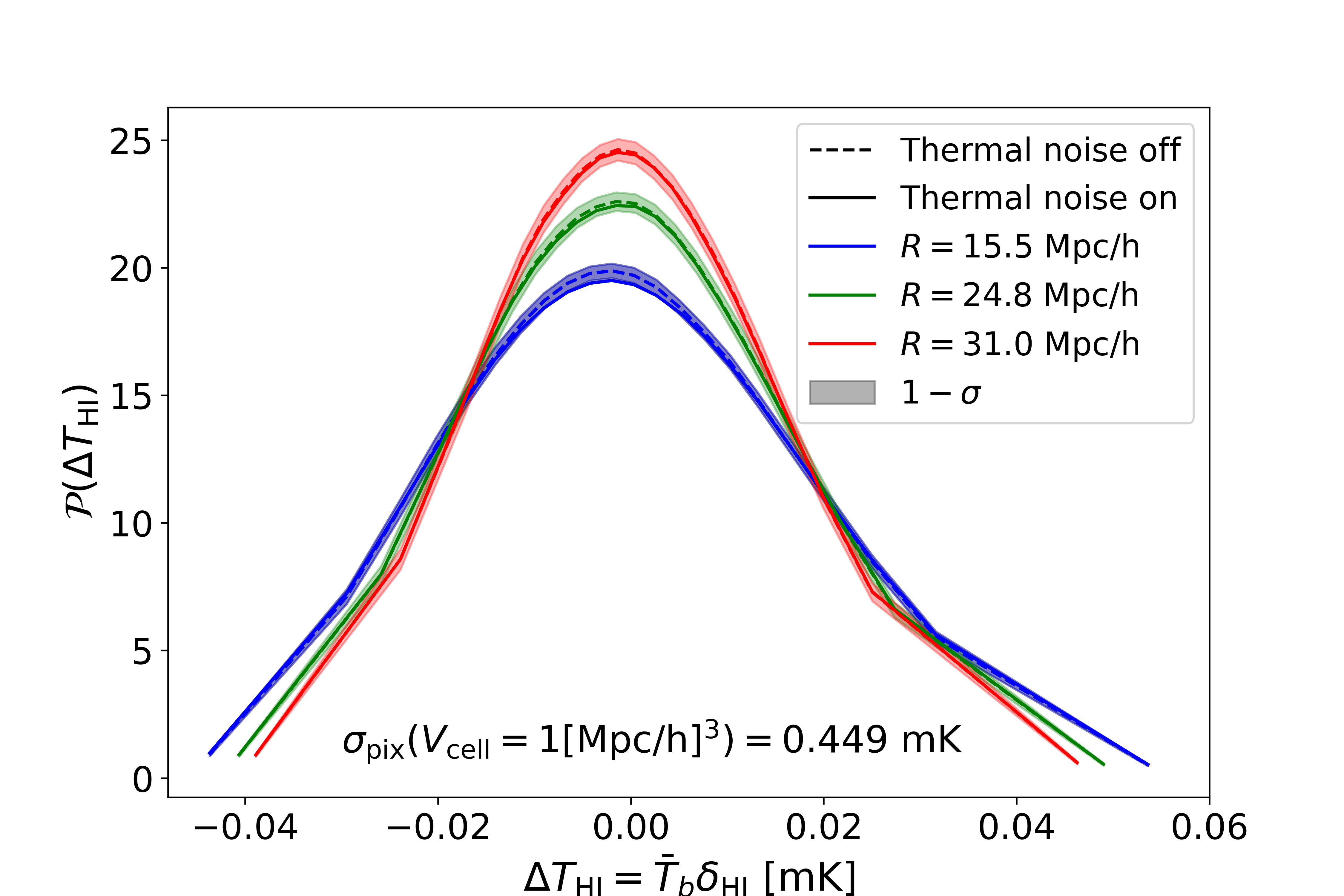}
        \caption{}
        \label{fig:Signal_and_noise}
    \end{subfigure}
    \hfill
    \begin{subfigure}[t]{0.37\textwidth}
        \centering
        \includegraphics[width=\textwidth]{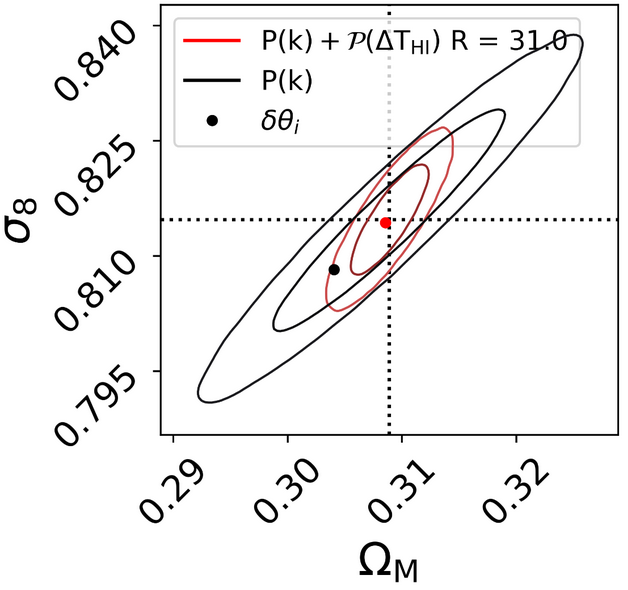}
        \caption{}
        \label{fig:CIC_Fisher}
    \end{subfigure}
    \caption{\textbf{Left:} Normalized HI PDF including telescope beam and foreground removal systematics, calculated for smoothing scales $R = 15.5,24.8,31.0$ Mpc/$h$. Dashed lines do not include thermal noise, and solid lines include thermal noise. $1-\sigma$ errors are represented as shaded contours. \textbf{Right:}  ($\Omega_M$, $\sigma_8$) Fisher forecast for SKA-Mid HI IM survey considering $z \sim 1.1-1.5$ and $f_{\rm sky} = 0.3$. We show 1$\sigma$ and 2$\sigma$ contours of the P(k) ($\rm{P(k)}$ + $\mathcal{P}(\Delta T_{\rm HI})$ at R = 31.0 $\rm{Mpc}/h$) as black (red) ellipses. Black and red points represent the cosmologies inferred from the predicted P(k) and PDF applied to the UNIT simulations, respectively.}
    \label{fig:CIC_BVG}
\end{figure}

\autoref{fig:Signal_and_noise} presents the Probability Density Function (PDF) of HI temperature fluctuations , $\mathcal{P}(\Delta T_{\rm HI})$, for three smoothing radii $R = 15.5, 24.8, 31.0$ Mpc/$h$ of simulated HI Intensity Maps where foreground-removal and telescope beam observational systematics have been taken into account. Our foreground removal technique approximately emulates a FastICA method with $N_{\rm IC} = 4$ independent components removed. Dashed and solid lines show the signal without and with including thermal noise, respectively, which slightly increases the variance. We also show the $1\sigma$ Jackknife errors as shaded regions. We consider a standard SKA-Mid AA4 configuration, although thermal noise is computed by modifying the pixel area and the channel width such that it corresponds to an Intensity Map with $L_{\rm cell} = 1$ Mpc/$h$. \autoref{fig:CIC_Fisher} presents a ($\Omega_M,\sigma_8$) Fisher forecast comparing the constraining power of the Power spectrum ($P(k)$), $(\sigma_{\Omega_M},\sigma_{\sigma_8})$ = (6.7,9.5) $\cdot 10^{-3}$, 
and its combination with the PDF at a single scale ($P(k)$ + $\mathcal{P}(\Delta T_{\rm HI})$ at $R = 31.0$ Mpc/$h$), $(\sigma_{\Omega_M},\sigma_{\sigma_8})$ = (2.8,5.2) $\cdot 10^{-3}$, for an SKA-Mid survey (with a redshift range $1.1 < z < 1.5$ and a sky fraction $f_{\rm sky} = 0.3$ corresponding to a survey volume of $V_{\rm survey} = 16.3$ $[\rm{Gpc}/h]^3$ ). 
Including the PDF alleviates the degeneracy between $\Omega_M$ and $\sigma_8$ when compared with the power spectrum \citep{vosgines2026}. The correlation factors for both parameter covariance matrices are $r_{\rm P(k)+\mathcal{P}(\Delta T_{\rm HI})} = $ $0.867$ and $r_{\rm P(k)} = $ $0.964$. Finally, we show that the cosmology inferred from the predicted statistics applied to the UNIT simulations is well below $1\sigma$ (black and red points), although higher-volume simulations are required as the survey volume considered in this setup exceeds the volume of our simulations. 

We use \textsc{hmcode-2020} \citep{Mead_2021} integrated in \textsc{CAMB} together with a bias and shot-noise prescriptions and observational systematic kernels to predict the HI power spectrum. We use \textsc{CosMomentum} \citep{Friedrich2020} to predict the HI PDF, considering bias and shot-noise prescriptions and introducing systematics as a variance rescaling. Those theoretical statistics are validated against statistics measured in high-resolution simulations at the fiducial cosmology (see Table 4 of \citet{Planck_2016}). We use all four available realizations of the high-resolution dark matter simulation \textsc{UNIT} \citep{Chuang_2019} snapshots at \( z = 1.321 \). For each realization, a semi-analytical model of galaxy evolution (SAGE) is applied to derive the cold gas mass distribution \citep{Croton_2016}. An HI prescription converts the cold gas mass to the HI mass, and telescope beam and foreground removal systematics are added to the HI catalog following \citet{SA_BVG_2021}. An Intensity Map is constructed in real space considering the assignment of a Nearest Grid Point (NGP) to cubic cells of side length $L_{\rm cell}=$1~Mpc$/h$ \citep{SA_BVG_2021}. Finally, we add thermal noise to each cubic cell, $\sigma_{\rm pix}$ \citep{Battye2013}. We compute the PDF of the HI field smoothed over spheres with three radii \( R = 15.5,\ 24.8,\ 31.0\ \mathrm{Mpc}/h \), considering 22 quantiles for each scale. 
We disregard the first and last PDF quantiles to limit the impact of nonlinearities and non-Gaussian distribution of the data. We consider the Power spectrum only up to $k_{\rm max} = 0.091$ $\rm{Mpc/h}$, since we expect smaller scales to be dominated by thermal noise. The data covariance matrix is computed for both $P(k)$ and its combination with the PDF at fixed cosmology, considering Jackknife resampling with $n=125$ realizations, and its inverse is corrected using the Hartlap factor \citep{kaufman1967, Hartlap}.

\begin{figure}
    \centering
    \includegraphics[width=0.6\textwidth]{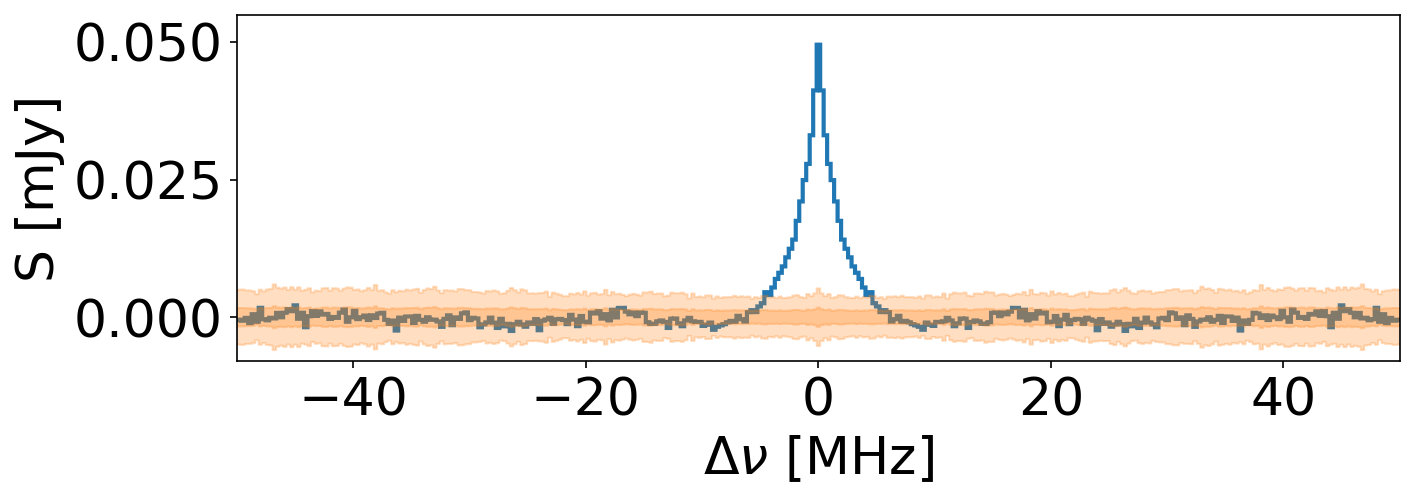}
    \caption{The forecast for the emission line stacking using single dish HI intensity mapping with SKA-Mid at $z\sim 1.1-1.5$. The blue solid line shows the expected HI signal. The outer orange shaded region shows the expected $3\sigma$ noise level, whereas the inner region shows the $1\sigma$ noise level.}
    \label{fig:stacksnr}
\end{figure}

Using the simulation of the dark matter halos and their HI content, we generate random realizations of the HI field to study the detectability of the stacking signal. Following the mock pipeline described in \cite{MeerKLASS:2025lmz}, we generate 100 realizations of the stacked 21-cm signal following the same noise statistics mentioned above. Assuming that halos with $M_{\rm HI}>10^{10.5}\,M_\odot$ will be selected for stacking, the stacked signal can be calculated using \autoref{eq:stack}. The realizations are then used to calculate the covariance of the stacked signal. The corresponding signal-to-noise ratio is shown in \autoref{fig:stacksnr}.


\subsection{21-cm bispectrum}
21-cm bispectrum is one of the promising higher-order statistics, which can be used to extract information contained in the non-Gaussian
components of the 21-cm signal. Measurements of the 21-cm bispectrum can independently constrain both linear and quadratic HI bias, thereby improving upon power-spectrum-only analyses. In a previous work, \cite{Sarkar_2019} studied the HI bispectrum using a set of semi-numerical simulations of the HI distribution. They determine the scale and redshift range where the HI bispectrum can be adequately modeled using the predictions of second-order perturbation theory, and they used this to predict the redshift evolution of the linear and quadratic HI bias. In the follow-up work, \cite{Chhabra_2025} showed that the HI mass-halo mass relation can be constrained by combining the 21-cm power spectrum and 21-cm bispectrum for all unique $k-\text{triangles}$. \cite{Cunnington_2021} presented a model of the redshift space 21-cm bispectrum based on second-order perturbation theory. Additionally, they studied the effect of the radio telescope beam and 21-cm foreground contamination.

\begin{figure*}
\centering
    \includegraphics[width=0.32\textwidth]{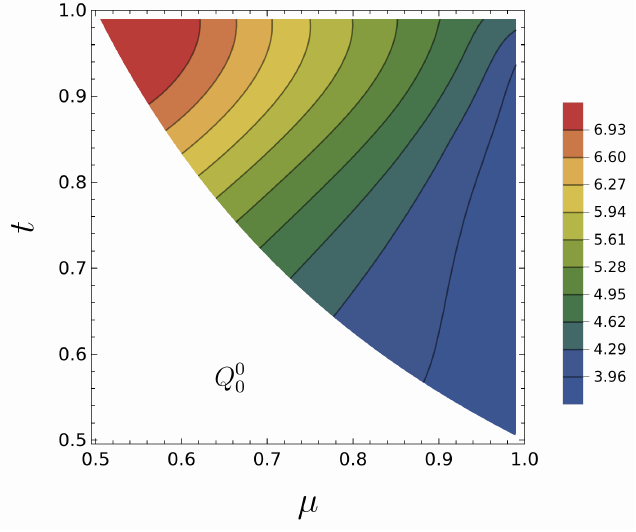}
    \hspace{\fill} 
    \includegraphics[width=0.32\textwidth]{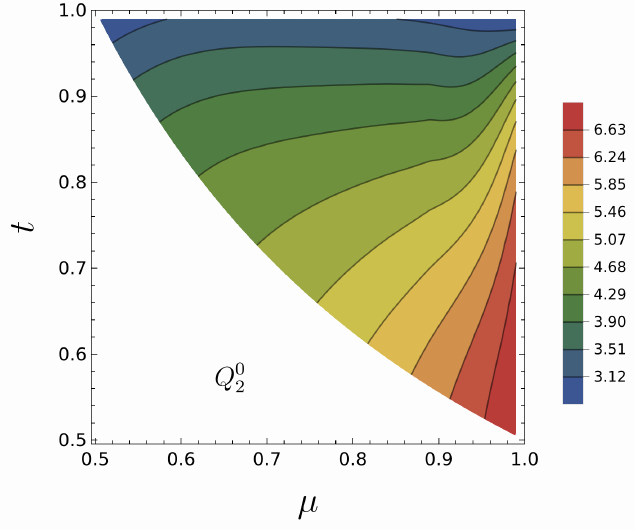}
    \hspace{\fill}
    \includegraphics[width=0.32\textwidth]{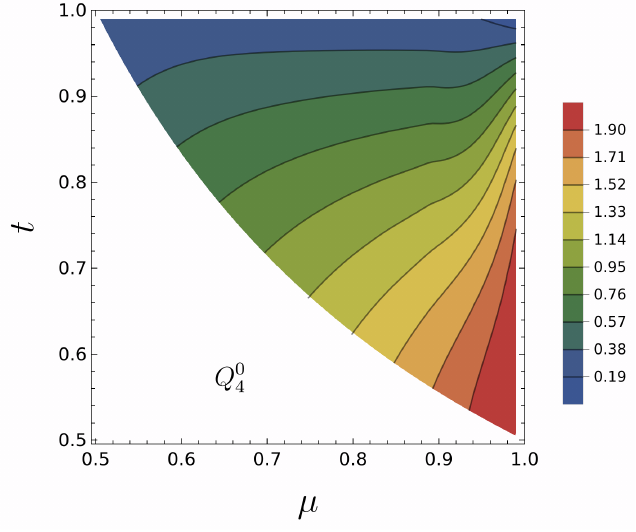}
    \vspace{0.5em} 
    \includegraphics[width=0.32\textwidth]{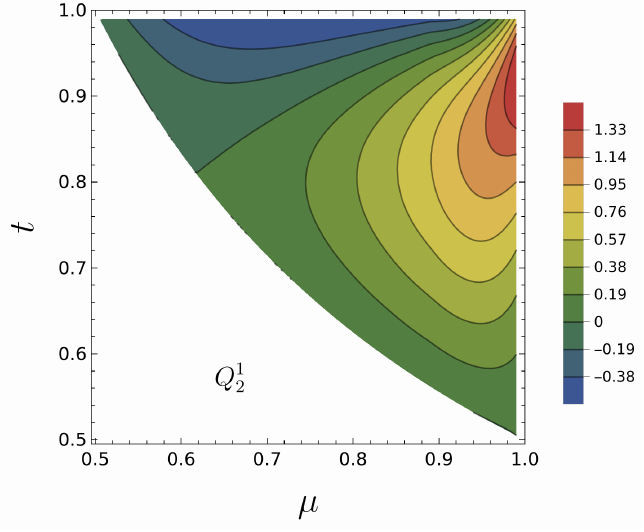}
    \hspace{\fill}
    \includegraphics[width=0.32\textwidth]{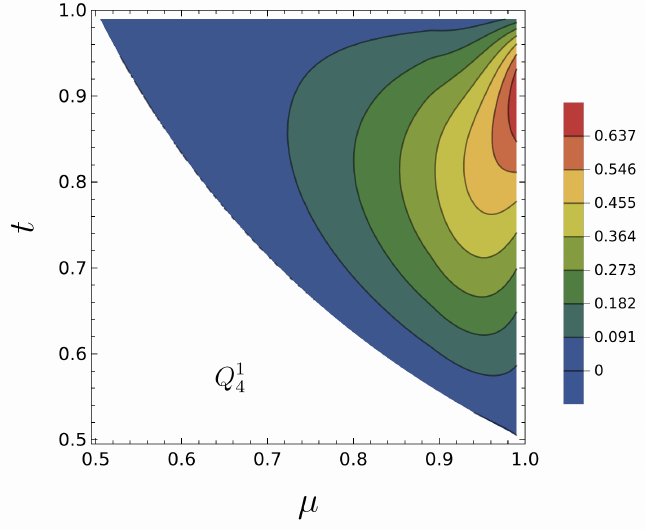}
    \hspace{\fill}
    \includegraphics[width=0.32\textwidth]{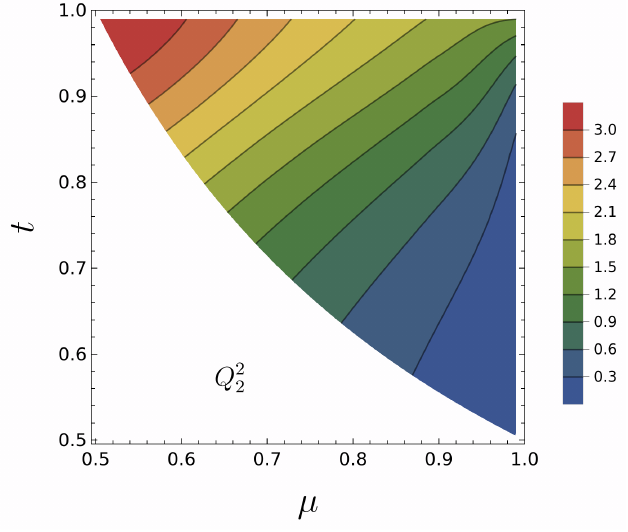}
    \caption{Heatmaps of the reduced bispectrum multipoles of the HI density contrast in the $(\mu, t)$-plane. The dimensionless reduced bispectrum multipoles are defined as $Q_\ell^m = B_\ell^m / B_{\rm HI}^r$, where $B_\ell^m$ denotes the redshift-space bispectrum multipoles including the FoG effect, and $B_{\rm HI}^r$ represents the real-space HI bispectrum. The results are shown for $k_1 = 0.1~ h{\rm Mpc}^{-1}$ at redshift $z = 2$. The redshift-space bispectrum multipoles $Q_\ell^0$ with $\ell = 0, 2, 4$ are amplified in the stretched configurations, while $Q_2^1$ and $Q_4^1$ exhibit enhancement along the linear triangles, peaking near the squeezed limit. In contrast, the $Q_2^2$ multipole shows the strongest enhancement near the equilateral configuration.}
    \label{fig:Q_multipoles}
\end{figure*}

The redshift space distortions can enhance the amplitude of the 21-cm bispectrum. Figure~\ref{fig:Q_multipoles} shows that at $z=2$ and $k_1=0.1\,h\,{\rm Mpc}^{-1}$, the monopole $Q_0^0$ has a strong overall enhancement in redshift space, with the largest amplification near the equilateral configuration and significant boosts toward stretched and squeezed triangles. Its peak location shifts with redshift, favoring stretched shapes at lower $z$. Among the $\ell=2$ quadrupoles, $Q_2^0$ is strongest for linear and stretched triangles ($\mu\!\rightarrow\!1$, $t\!\lesssim\!0.8$) and becomes suppressed near the equilateral limit.  
The $Q_2^1$ mode alternates between enhancement and suppression across the $(\mu,t)$ plane, peaking near squeezed configurations and turning negative near the L-isosceles limit. The $Q_2^2$ mode mirrors the monopole behavior, with its largest amplitude near the equilateral shape. Higher-order multipoles ($\ell\!\ge\!4$) exhibit more complex but weaker angular structure; FoG damping suppresses them more effectively, leaving only modest enhancement in $Q_4^0$ for stretched configurations and mild signatures in $Q_4^1$ near squeezed triangles.

\subsubsection{Detectability of the 21-cm bispectrum}
\begin{figure}[htbp]
    \centering
    \includegraphics[width=0.9\textwidth]{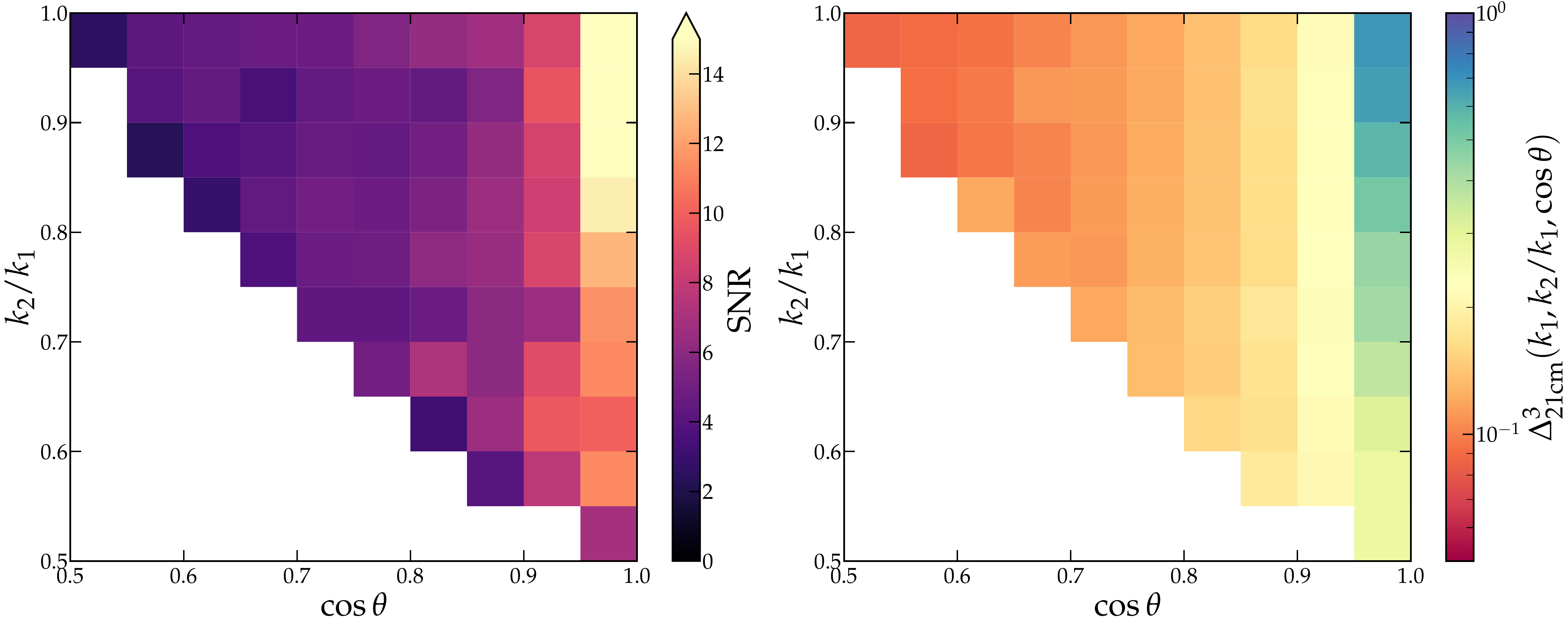}
    \caption{\textbf{Right:} 21-cm bispectrum for all unique $k-\text{triangles}$ for $k_1 = 0.61 \text{Mpc}^{-1}$ at $z=0.99$. \textbf{Left:} The signal-to-ratio of 21-cm bispectrum for all unique $k-\text{triangles}$ for 300 hours of observation per pointing with SKA-Mid.
    \label{fig:bs_SNR_GAEA_z099}}
\end{figure}

In the \autoref{fig:bs_SNR_GAEA_z099}, we show the forecast for the detectability of spherically averaged 21-cm bispectrum for all unique $k-\text{triangles}$ with SKA-Mid observations. The right panel of the \autoref{fig:bs_SNR_GAEA_z099} show the 21-cm bispectrum for all unique $k-\text{triangles}$ for $k_1 = 0.61 \text{Mpc}^{-1}$ at redshift $z=0.99$. The 21-cm intensity maps used for this forecast were generated by postprocessing the galaxy catalogs from the GAEA semi-analytic galaxy formation model (\citealp[and references therein]{DeLucia_2014, Hirschmann_2016, DeLucia_2024, Fontanot_2025}).
 For the signal-to-noise ratio (SNR) estimation, we only include the uncertainty due to thermal noise and neglect any contribution from the cosmic variance~\citep{Floss_2023} and foreground residuals~\citep{Cunnington_2021}.
 To estimate the SNR, we simulate 50 statistically independent realizations of the system noise map within the volume of the 21-cm signal, and add the system noise and 21-cm signal. To estimate the system noise map, we considered 300 hours of SKA-Mid AA4 observations for each pointing in the interferometric mode. The right panel of \autoref{fig:bs_SNR_GAEA_z099} shows the SNR prediction of the 21-cm bispectrum for all unique $k-\text{triangles}$ for $k_1 = 0.61 \text{Mpc}^{-1}$ at redshift $z=0.99$. Except for a few unique $k-\text{triangles}$, all others have SNR higher than $3\sigma$ and $k-\text{triangles}$ in the vicinity of the squeezed-limit ($k_2/k_1 = 1$ and $\cos\theta \rightarrow 1$) show SNR higher than $10\sigma$~\citep{Noble:2026oqm}.
 
\subsubsection{Detectability of redshift-space 21-cm bispectrum multipoles}
To assess detectability, we model the thermal noise for SKA-Mid AA4 with $N_{\rm ant}=197$ antennas of diameter $D_{\rm dish}=15\,{\rm m}$ and aperture efficiency $\eta=0.7$. The effective area per dish is $A_e=\eta\pi D_{\rm dish}^2/4$. Each Fourier mode corresponds to a baseline $b=(\lambda k_\perp r)/(2\pi)$, where $\lambda=0.21(1+z)$\,m and $r(z)$ is the comoving distance. Assuming a baseline distribution $\rho(b)\propto b^{-2}$, an observing time $T_{\rm obs}=400$\,hr per pointing, bandwidth $B=32$\,MHz, and system temperature $T_{\rm sys}=60$\,K at $z=2$, the thermal-noise power is
\begin{equation}
P_N(k,\mu) = \frac{A_e\,T_{\rm sys}^2\,r^2(z)}{t_k}\,\frac{dr}{d\nu},
\end{equation}
where $t_k$ is the effective integration time per mode.  
The bispectrum multipoles are then estimated as weighted sums over closed triangle triplets within each bin~\citep{Mazumdar:2020bkm, Pal:2025hpl, Pal:2025zep}.  
The covariance matrix $C_{\ell\ell'}^{mm'}(k_1,\mu,t)$ includes both cosmic variance and thermal noise contributions.
The signal-to-noise ratio (SNR) for each multipole is
\begin{equation}
{\rm [SNR]}_\ell^m(k_1,\mu,t)
= \frac{B_\ell^m(k_1,\mu,t)}{\sqrt{C_{\ell\ell}^{mm}(k_1,\mu,t)}}.
\end{equation}

\begin{figure*}
\centering
    \includegraphics[width=0.32\textwidth]{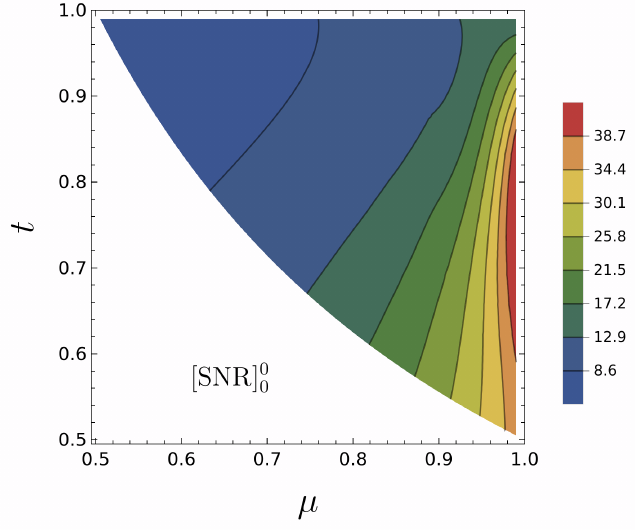}
    \hspace{\fill}
    \includegraphics[width=0.32\textwidth]{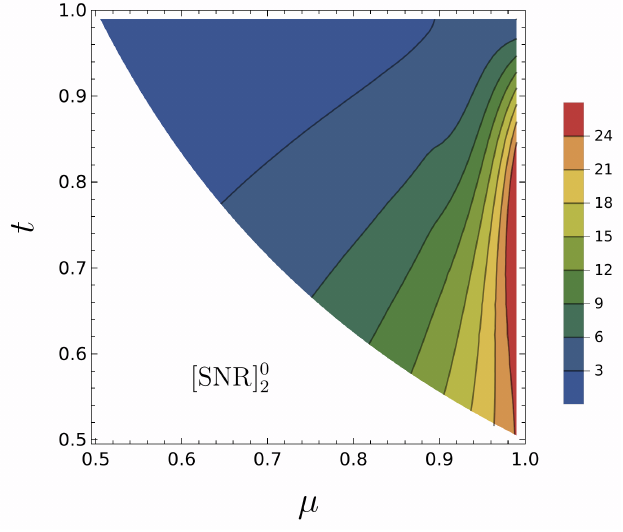}
    \hspace{\fill}
    \includegraphics[width=0.32\textwidth]{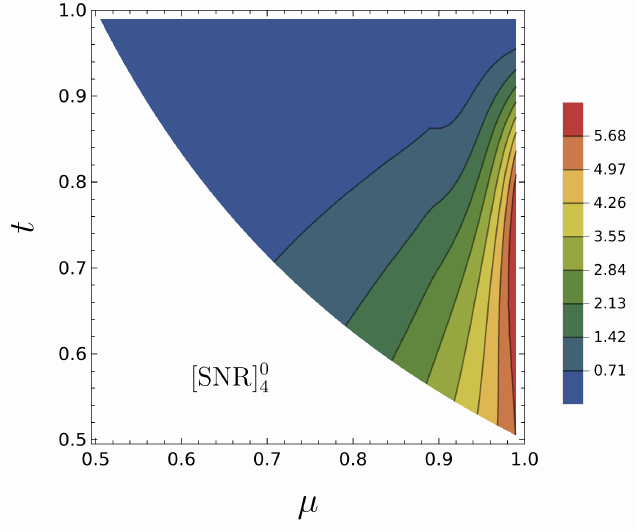}
    \vspace{0.5em}
    \includegraphics[width=0.32\textwidth]{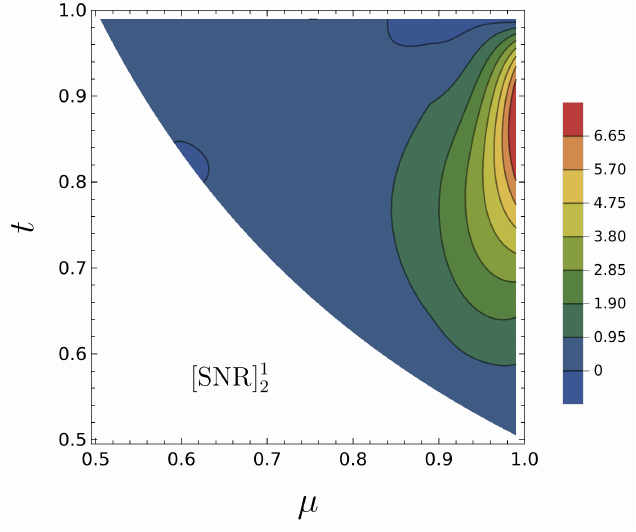}
    \hspace{\fill}
    \includegraphics[width=0.32\textwidth]{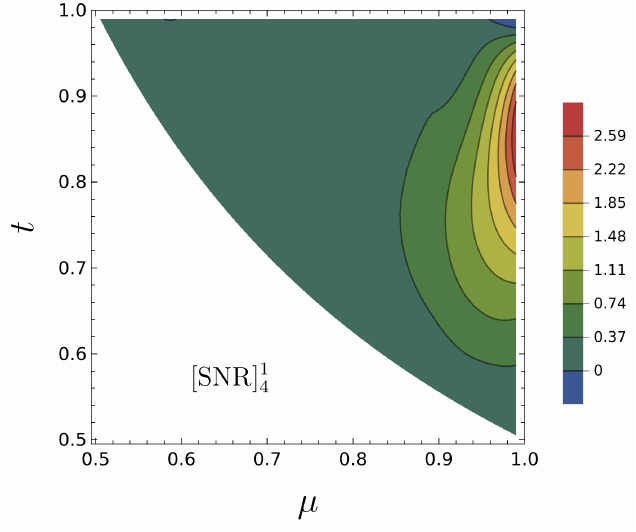}
    \hspace{\fill}
    \includegraphics[width=0.32\textwidth]{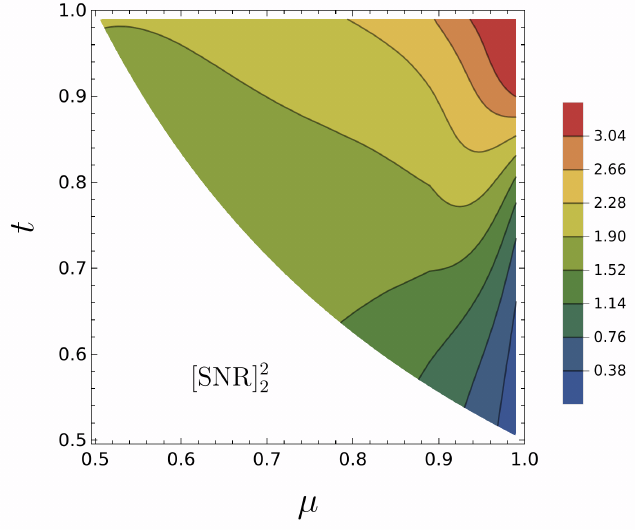}

\caption{Signal-to-noise ratio (SNR) maps of the bispectrum multipoles up to $(\ell, m) = (4, 1)$ in the $(\mu, t)$-plane. The SNR is computed for $k_1 = 0.1\, h\, {\rm Mpc}^{-1}$ at redshift $z = 2$. For the $B_{\ell}^0$ multipoles with $\ell = 0, 2, 4$, the SNR for SKA-Mid peaks at $\mu \rightarrow 1$ and $0.6 \leq t \leq 0.85$. The $B_2^1$ and $B_4^1$ multipoles exhibit maximum SNR near the linear triangle configurations (i.e., $\mu \rightarrow 1$, $0.75 \leq t \leq 0.95$), while the $B_2^2$ multipole attains its highest SNR near the squeezed limit ($\mu \rightarrow 1, t \rightarrow 1$).}
\label{fig:snr_multipoles}
\end{figure*}

Using the noise and covariance model above, we compute the SNR maps for SKA-Mid at $z=2$ and show the results in Figure~\ref{fig:snr_multipoles}.  
The SNR peaks in the limit of linear-triangles ($\mu\!\rightarrow\!1$, $0.6\!\lesssim\!t\!\lesssim\!0.85$), where FoG suppression is weakest, and declines toward equilateral configurations. For the monopole ($\ell=0,m=0$), the maximum SNR reaches $\sim40$ near linear triangles and $\sim8$ near equilateral. The quadrupole ($\ell=2,m=0$) attains $\sim25$ in the same region, decreasing to $\sim3$ near equilateral.  
The hexadecapole ($\ell=4,m=0$) peaks around $\sim6$ near linear configurations but is heavily damped at equilateral.  
Among the azimuthal modes, $(\ell,m)=(2,1)$ peaks near the linear limit with SNR $\lesssim10$, $(4,1)$ remains small (maximum $\sim2.5$), and $(2,2)$ shows a broader pattern peaking toward the squeezed limit with SNR $\sim3$.  
Overall, the monopole dominates the detectable signal, followed by the $\ell=2$ modes, with higher $\ell$ and $m>0$ multipoles measurable but subdominant. Configurations close to linear triangles offer the highest detectability, whereas equilateral shapes are most strongly affected by FoG damping. These results highlight the power of the multipole framework to isolate anisotropic signatures of RSD and to pinpoint the most informative regions of configuration space for future 21-cm surveys with SKA-Mid~\citep{cosmicVisions21cm:2018rfq, Dash:2024jfi, Yadav:2024kzh, Dash:2023scq, Mazumdar:2020bkm, Pal:2025hpl, Pal:2025zep}.

\subsection{Marked Power Spectrum}
In this section, we take a step further and explore how marked statistics can be leveraged in the context of the upcoming 21-cm intensity mapping surveys with SKA-Mid. Given its sensitivity and redshift coverage spanning $z \sim 0.2$ to $z \sim 3$, SKA-Mid will enable precise measurements of large-scale HI clustering during the post-reionization era. This makes it an ideal testbed for applying marked power spectrum techniques to observationally motivated scenarios. Our goal here is to demonstrate the application of density-dependent weighting schemes to the HI field, but now within realistic SKA-Mid forecast simulations that include instrumental effects such as beam smoothing and thermal noise. 

To perform this exercise, we use a simple functional form of the mark first used in \cite{White_2016}, as

\begin{equation} \label{Mark:W16}
     m(\mathbf{x},t;R, \rho_{\star},p) = \Bigg[ \frac{\rho_\star + 1}{\rho_\star + \rho_R(\mathbf{x},t)} \Bigg]^p
\end{equation}
where the free parameters $R$, $\rho_\star$, and $p$ are referred to as the mark parameters. Both $\rho_R$ and $\rho_*$ are expressed in units of the mean density $\bar{\rho}$. The parameter $\rho_R$ quantifies the sensitivity of the mark to the local density around each point. The exponent $p$ introduces an additional contrast between points residing in different environments by weighting them differently.

\begin{figure}[htbp]
    \centering
    \includegraphics[width=\linewidth]{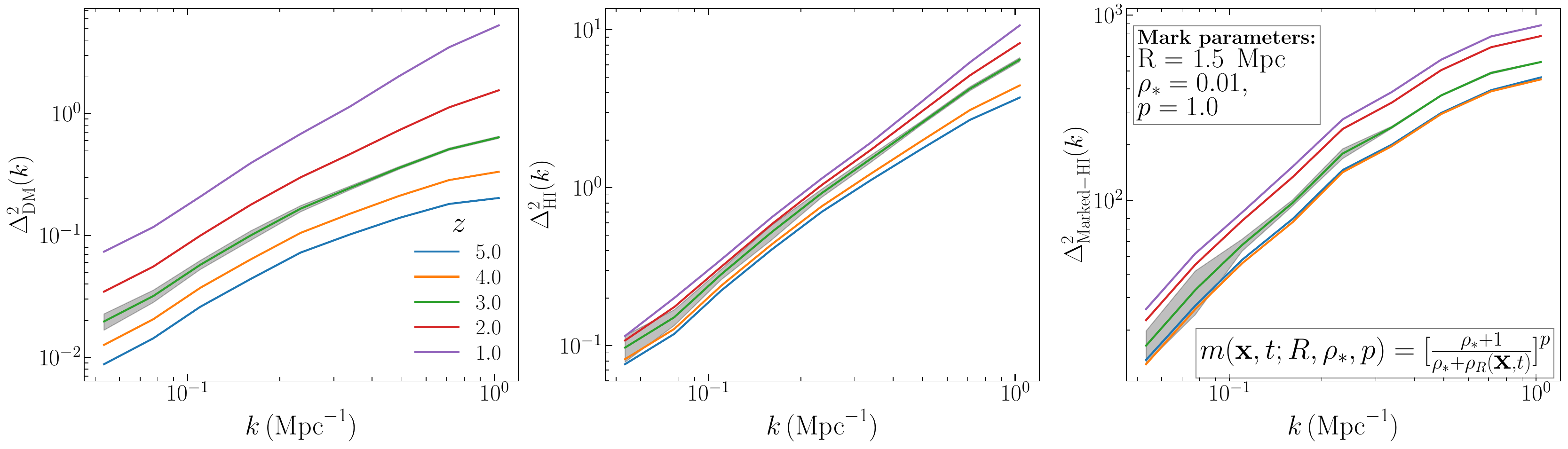}
    \caption{The dimensionless power spectrum of the matter (left panel), the H\,\textsc{i} (middle panel), and the H\,\textsc{i} with density values $\rho/\bar{\rho} < 3.0$ (right panel). The distribution is smoothed for a fixed radius of $R = 280 \, \mathrm{kpc}$. The power spectra are demonstrated as a function of $k$-mode at five different redshifts in the range $z \sim 1$--$5$. The shaded regions in each panel represent $\pm 1\sigma$ spread around the mean value at $z = 3$.}
    \label{fig:mark_model_HOD5_v2.pdf}
\end{figure}

\begin{figure}[htbp]
    \centering
    \includegraphics[width=\linewidth]{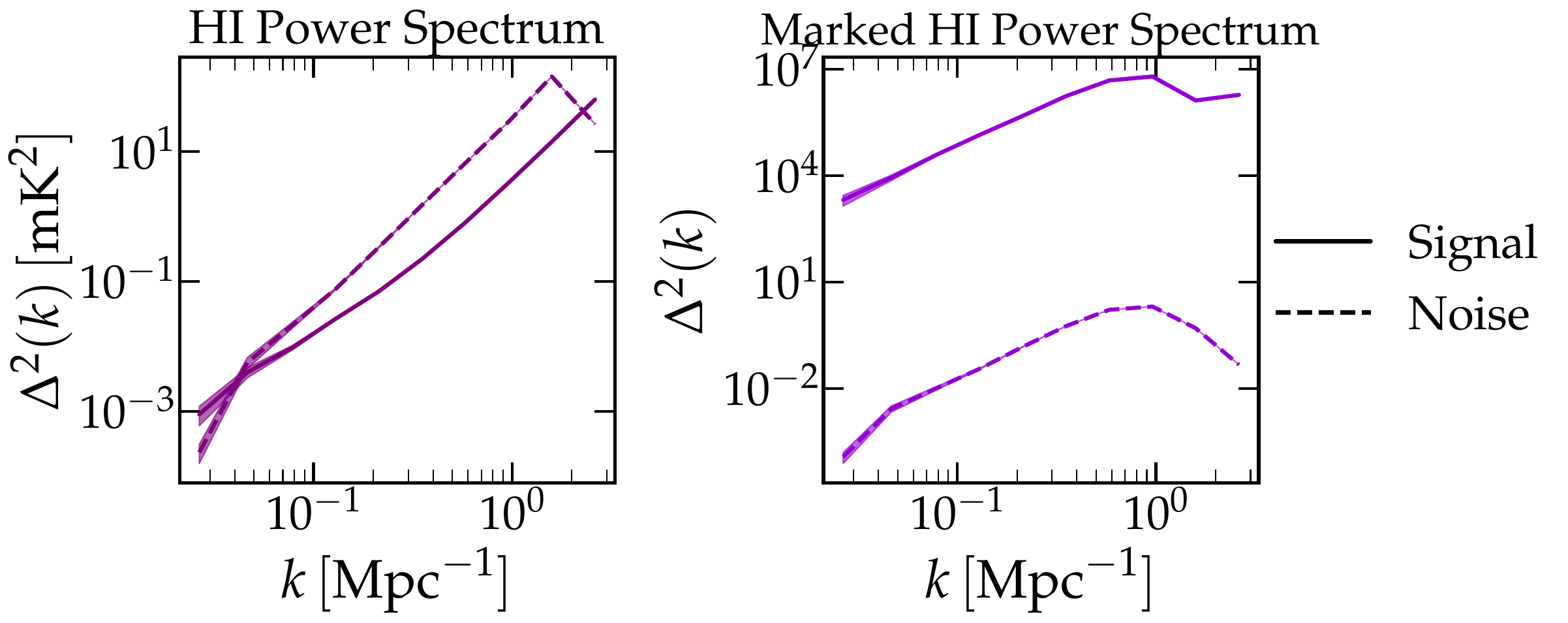}
    \caption{Comparing the signal strength, at different $k$-modes, against the corresponding noise levels using marked statistics designed to boost the signal strength at redshift $z=1$ with the standard signal power spectrum. \textit{Left panel}: The dimensionless HI power spectrum and mark power spectrum of the signal fluctuation and noise. \textit{Right panel}: The dimensionless HI power spectrum and marked field power spectrum of the signal and noise. The shaded region corresponds to $\pm1\sigma$ sigma spread around the mean.}
    \label{fig:Mark_SNR_HI.pdf}
\end{figure}

As mentioned earlier, suitable marks can selectively highlight distinct environments in the map. It is possible to construct a functional form of the mark that boosts the signal strength. To validate this, we simulate mock LIM noise maps of [HI] 21-cm emission lines at $z = 1$, for SKA-Mid AA4 configuration with 5000 hours of integration time. Modeling of the [HI] 21-cm signal is done by postprocessing the IllustrisTNG simulation subhalo catalog as done in \citep{dosibhatla_2025_lss-morphology} using an analogous functional form of the mark as given in Figure \ref{fig:mark_model_HOD5_v2.pdf} for the HI 21-cm signal. To boost the signal strength, we fix the mark parameters to up-weigh the high density points in the maps at a relatively small smoothing length scale of $R=3 \, {\rm Mpc}$. In Figure \ref{fig:Mark_SNR_HI.pdf}, we compare the strength of signal fluctuation as a function using the marked power spectrum with their corresponding noise power spectrum. The signal and the noise are marked and up-weighted in the same way. As clearly observed, a suitable mark is able to boost the signal strength significantly, which is evident from the marked power spectrum as compared to the standard power spectrum.
 
\subsection{Three-point correlations and beyond}


The Alcock–Paczynski (AP) effect imprints anisotropic geometric distortions on higher-order statistics by differentially rescaling transverse and radial Fourier components when an incorrect distance–redshift mapping is assumed. This anisotropic mapping reshapes triangle (bispectrum) and tetrahedral (trispectrum) configurations in $k$-space, altering both the amplitudes and angular dependences of measured bispectra and trispectra and producing multipole mixing that must be modeled to avoid biased cosmological inferences \citep{Karagiannis:2022ylq, Philcox:2022frc, Randrianjanahary:2023rgp}. For the bispectrum, the AP distortion rescales the three triangle sides and mixes bispectrum multipoles; robust inference therefore requires joint modeling of AP, redshift-space distortions, bias, and survey effects (beam, window, and foreground mode loss) so that geometric information [e.g. $D_A(z)$, $H(z)$ or the AP scaling parameters $\alpha_{\parallel} \alpha_{\perp}$] can be disentangled from systematic contributions. For the trispectrum, the situation is yet more complex—the richer set of shape degrees of freedom (quadrilaterals) means AP both rescales and can change shape classification, and explicit trispectrum with AP treatments remain comparatively sparse in the 21-cm intensity-mapping literature. The 21-cm experiments, such as SKA-Mid, therefore need to include AP transformations into their higher-order statistics analysis pipeline to extract geometric constraints while marginalizing over astrophysical and instrumental nuisance parameters. 

In this context, both primordial (PNG) and induced sources of non-Gaussianity are important. From the primordial side, non-zero bispectra and trispectra can arise from inflationary physics, and the 21-cm signal at large scales can retain imprints of such early-Universe signatures. For example, the local-type PNG generates a characteristic scale-dependent bias in the bispectrum that can be isolated with appropriate triangle configurations \citep{karagiannis_2018_PNG-forecasts}. On the other hand, induced non-Gaussianity arises from late-time structure formation, contributes significantly to the bispectrum and trispectrum at intermediate and small scales. SKA-Mid is expected to shed light on both of them. 




Beyond higher-order statistics such as bispectrum and trispectra, modern machine learning offers the opportunity to learn statistics that capture non-Gaussian features, and to do so directly from 21-cm maps. Here, a compressor learns a summary that is optimal in the sense of providing unbiased and precise posteriors for a given inference task. Optionally, the summary can be Gaussianised to more closely resemble power spectra as summary statistics. The performance of the learned statistics can then be judged, for example, via simulation-based calibration of posteriors, or measurements of mutual information~\citep{Makinen2024}. This approach has been shown, for example, to infer properties on inflation and dark matter from 21-cm maps, and provide more robust results than fixed summaries when faced with noise and modeling uncertainties~\citep{Ore2025, Schosser2025, Andrianomena2025}. Such methods can certainly be explored with SKA-Mid-produced 21-cm maps.

\subsection{Starlet $\ell_1$-norm}

\begin{figure}[htbp]
    \centering
   \begin{subfigure}{0.54\textwidth}
        \centering
        \includegraphics[width=\textwidth]{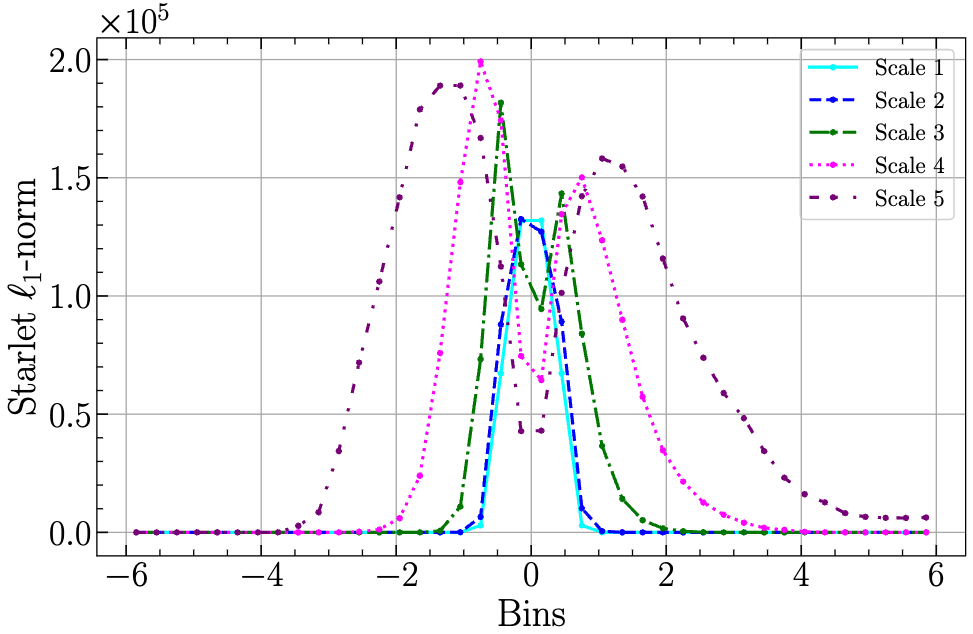}
        \caption{}
        \label{fig:l1}
    \end{subfigure}
    \hfill
    \begin{subfigure}{0.41\textwidth}
        \centering
        \includegraphics[width=\textwidth]{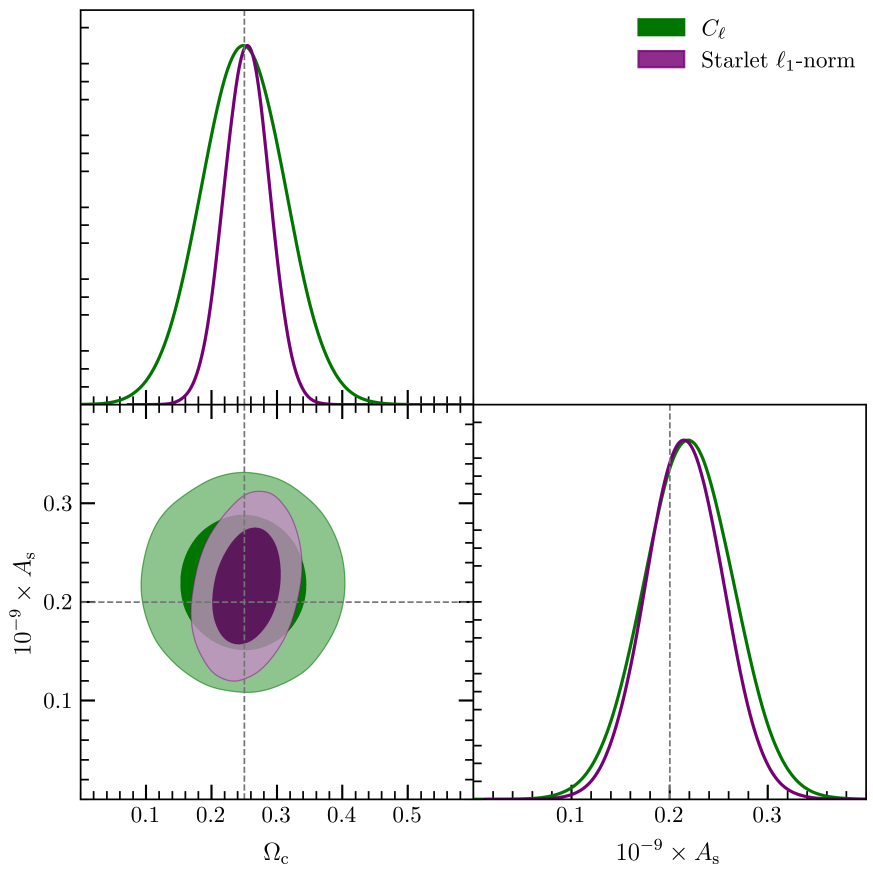}
        \caption{}
        \label{fig:l1_contour}
    \end{subfigure}
    \hfill
 
    \caption{\textbf{Left:} $\ell_1$-norm of the starlet coefficients as a function of the bin threshold for five decomposition scales of the HI intensity map. \textbf{Right:}  Posterior distributions for $\Omega_{\mathrm{c}}$ and 
$10^{-9}\,A_{\mathrm{s}}$ inferred with simulation-based inference (SBI). 
The green contours and curves correspond to the power spectrum $C_\ell$, 
and the purple ones to the Starlet $\ell_1$-norm. Contours indicate the 
$68\%$ and $95\%$ confidence regions. The inference was performed on three 
parameters $(\Omega_{\mathrm{c}},\,10^{-9}A_{\mathrm{s}},\,h)$, of which only 
two are displayed, as $h$ is weakly constrained \citep{Gorbatchev:2026zyj}.}
    \label{fig:l1_norm}
\end{figure}

Figure \ref{fig:l1_norm} shows the $\ell_{1}$-norm statistic of the starlet coefficients as a function of the bin threshold for five decomposition scales of the HI intensity map (left), alongside the posterior distributions for $\Omega_{\mathrm{c}}$ and $10^{-9}A_{\mathrm{s}}$ inferred via simulation-based inference (right). The green and purple contours correspond to the angular power spectrum $C_\ell$ and the starlet $\ell_{1}$-norm statistic, respectively, with shaded regions indicating the 68\% and 95\% confidence intervals. Inference is performed on the three parameters $(\Omega_{\mathrm{c}},10^{-9}A_{\mathrm{s}},h)$, of which only the first two are shown, as $h$ is weakly constrained \citep{Gorbatchev:2026zyj}.

The starlet $\ell_{1}$-norm statistic produces significantly tighter posterior contours than the angular power spectrum, highlighting its enhanced sensitivity to cosmological information. This improvement arises from its multiscale, higher-order nature, which captures non-Gaussian features of the intensity field that are inaccessible to two-point statistics alone.

The results are based on lognormal simulations generated with GLASS \citep{GLASS} and CAMB (\href{https://camb.info}{https://camb.info}), analyzed using a likelihood-free SBI pipeline implemented in JAXILI (\href{https://jaxili.readthedocs.io/en/latest/}{https://jaxili.readthedocs.io/en/latest/}), which efficiently handles the intractability of the likelihood of the starlet $\ell_{1}$-norm statistic. The maps correspond to full-sky simulations with a sky fraction of $f_{\mathrm{sky}} = 0.19$, defined by the applied mask, at a resolution of $N_{\mathrm{side}} = 1024$ and at redshift $z=0.425$. They include instrumental noise and are smoothed with a 30 arcmin Gaussian beam , and do not include any foreground contaminants. The starlet decomposition is performed over five scales with effective sizes of [3.7, 5.6, 11, 22, 60] arcmin.

\subsection{Morphological Measures in the Image Domain}

Imaging line intensity maps at sub-Mpc resolutions can reveal the intricate cosmic web structure of the universe. Summary statistics in the image domain can be computed on these maps to extract the non-Gaussian and morphological information from these fields. These summary statistics can, in principle, tighten constraints and break degeneracies of astrophysical as well as cosmological parameters due to their higher information content. 

\begin{figure}[htbp]
    \centering
    \begin{subfigure}[t]{0.43\textwidth}
        \centering
        \includegraphics[width=\textwidth]{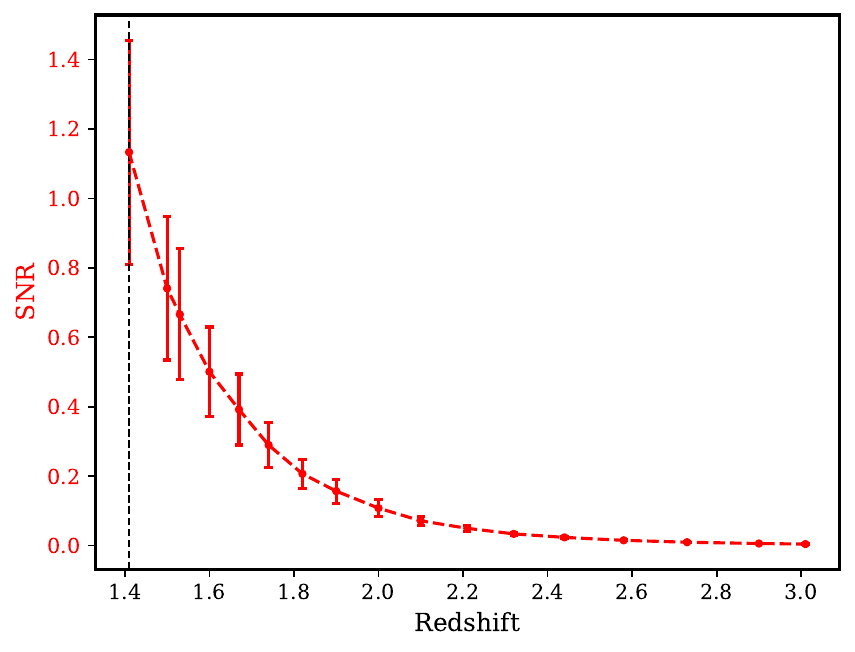}
    \end{subfigure}
    \hfill
    \begin{subfigure}[t]{0.55\textwidth}
        \centering
        \includegraphics[width=\textwidth]{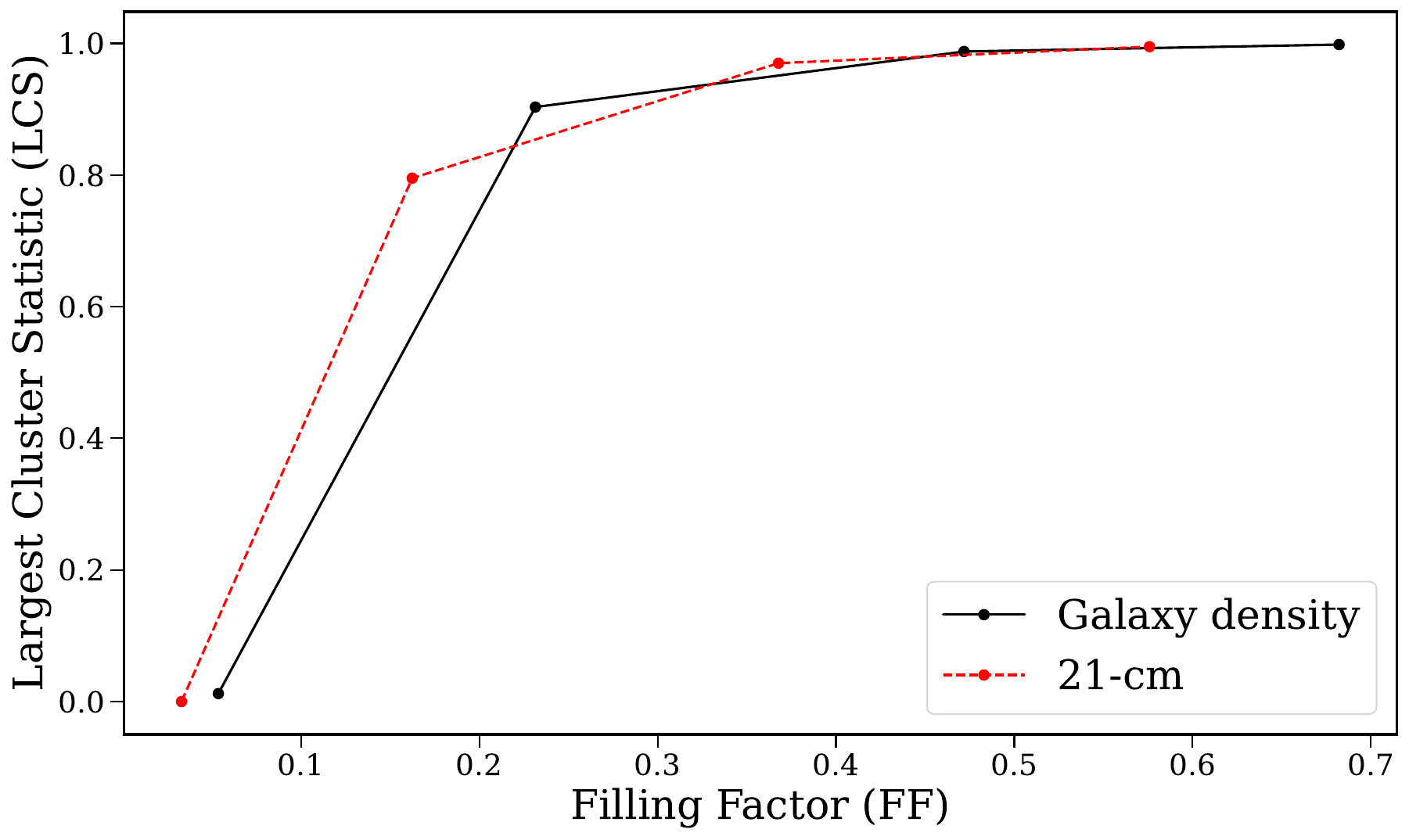}
    \end{subfigure}
    \caption{\textbf{Left:} Signal-to-noise ratios (SNR) of the 21-cm brightness temperature fluctuations at spatial resolution $\delta x \simeq 1$ Mpc in the overlapping redshift range where SKA-Mid and SKA Phase 2 can observe the 21-cm and CO(1-0) LIM signals, respectively. The noise estimates are made assuming Gaussian random thermal noise for 5000 hours per pointing of SKA-Mid observations. The black dashed line corresponds to $z=1.41$, the redshift where the 21-cm maps just become signal-dominant. The $1\sigma$ error bars denote standard deviation in SNR across $8 \, (150 \, {\rm Mpc})^3$ subcubes each corresponding to a $4 \, {\rm deg}^2$ SKA-Mid survey, due to cosmic variance in signal fluctuations. \textbf{Right:} Percolation curves for the galaxy mass density (solid black) and 21-cm brightness temperature (dashed red) maps at a spatial resolution of $0.5$ Mpc.}
    \label{fig:SNR_and_percolation}
\end{figure}

However, obtaining maps at such a high spatial and hence angular resolution is a significant challenge due to foreground contamination coupled with instrumental systematics. Instrumental noise also washes out the information of the LSS in intensity maps. Imaging at high resolution requires information from long baselines, which are sparsely distributed. Besides, a high line-of-sight resolution requires a high spectral resolution, which in turn boosts the thermal noise contamination. In this section, we refer to the SKA-Mid AA4 setup with frequency coverage from $0.58$ GHz to $15$ GHz as SKA-Mid, and the same array with frequency coverage extended to $50$ GHz as SKA Phase 2. The left panel of Figure \ref{fig:SNR_and_percolation} shows the SNR estimates for mock 21-cm maps observed by SKA-Mid in the redshift range where SKA Phase 2 can observe the CO(1-0) line \citep{dosibhatla_2025_lss-morphology, Sarkar01.2026.SKA} in the presence of thermal noise only. In this section, the signal maps are modeled by postprocessing IllustrisTNG simulation subhalo catalogs as per \cite{dosibhatla_2025_lss-morphology}.


It is seen that even with 5000 hours of observation time at a channel-averaged spectral resolution of eight SKA-Mid channels, the 21-cm maps are highly noise-dominated. In this chapter, image analysis results are shown at $z = 1.41$, where the 21-cm SNR approaches $1$.

\subsubsection{Largest Cluster Statistic}

The value of FF at which percolation occurs is expected to differ for different maps. The analysis of \cite{Bharadwaj_2000} implies that fields with a higher degree of overall filamentarity percolate at lower filling factors. Thus, percolation analysis can distinguish between astrophysical or cosmological models that induce different degrees of filamentarity in an intensity map.

The right panel of Figure \ref{fig:SNR_and_percolation} shows the percolation curves for the galaxy density field and 21-cm intensity maps at $z=1.41$, simulated from the same underlying galaxy distribution. The FF and LCS values are computed after every iteration of the coarse-graining scheme mentioned in section \ref{subsubsec:LCS_formalism}. As the two maps follow different percolation curves despite being tracers of the same galaxy distribution, the LCS is sensitive to the complex astrophysics of line emissions. For more details, see \cite{dosibhatla_2025_lss-morphology}. Due to its sensitivity to factors that affect the connectivity and filamentarity of the LSS, LCS can be an effective higher-order summary statistic to infer both astrophysics and cosmology.
 


\subsubsection{Local Dimension}

\begin{figure}[htbp]
    \centering
    \includegraphics[width=0.55\linewidth]{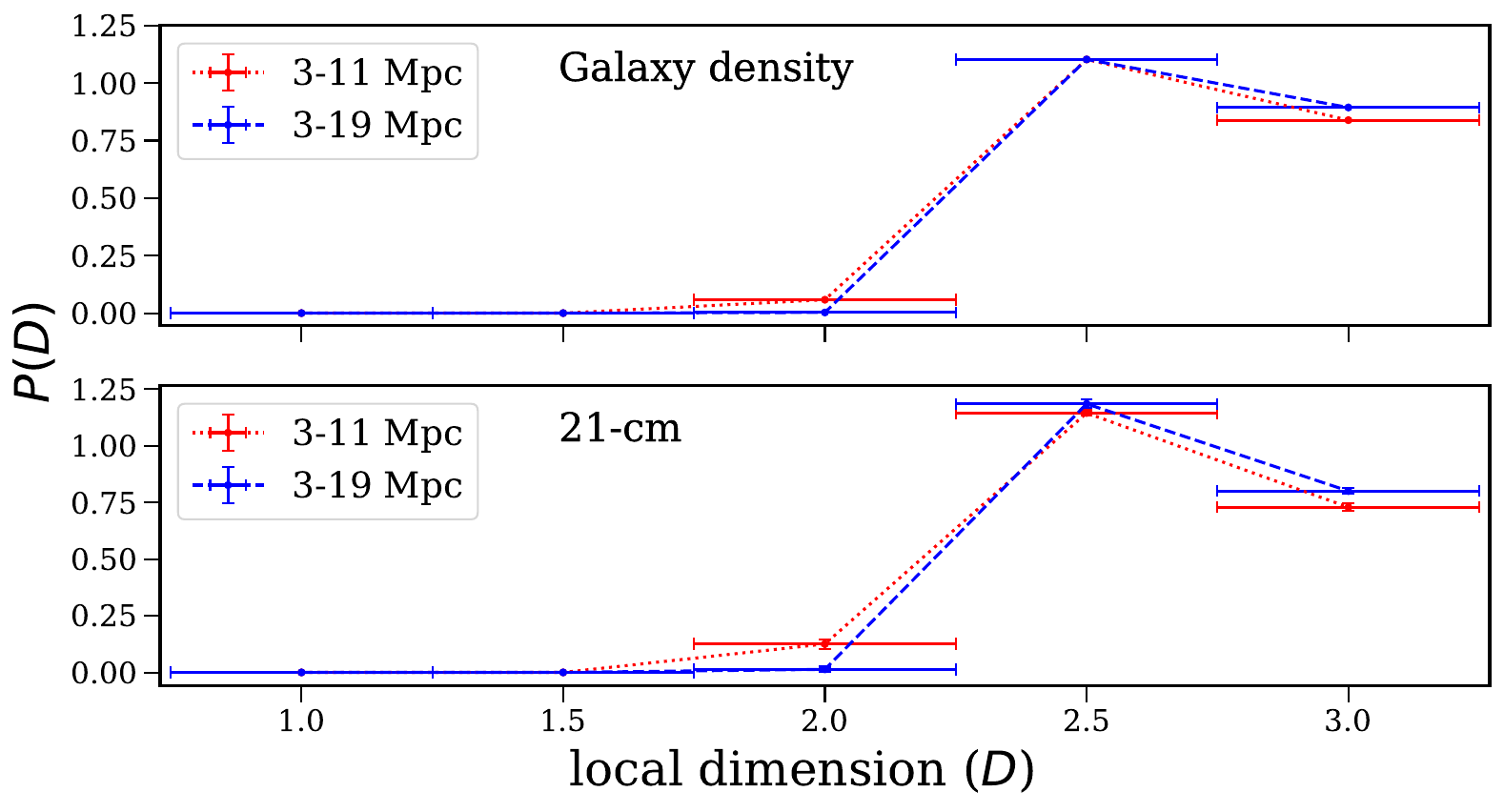}
    \caption{Distribution of local dimensions of classifiable cells out of $10^5$ randomly sampled bright cells in the galaxy density (top) and 21-cm brightness temperature (bottom) maps simulated with galaxy positions in redshift space. The $D$ values are binned into intervals of size 0.5. The different curves correspond to different length scales specified by the $R_{\rm min}$ and $R_{\rm max}$ values. The vertical $1\sigma$ error bars denote standard deviation in $P(D)$ across $8 \, (150 \, {\rm Mpc})^3$ 21-cm subcubes, each corresponding to a $4 \, {\rm deg}^2$ SKA-Mid survey for $10^4$ randomly sampled bright cells.}
    \label{fig:locdim_signal_RSD}
\end{figure}

Once the local dimensions of a subset of cells in an intensity map are determined as in \ref{subsubsec:local_dimension_formalism}, the fraction of cells lying in filaments, sheets, and volume-filling environments (nodes and voids) can be computed. Figure \ref{fig:locdim_signal_RSD} shows the distribution of local dimensions of the galaxy density field and 21-cm intensity maps divided into bins of width $0.5$. The maps have a spatial resolution of $\delta s = 0.87$ Mpc, corresponding to a spectral resolution of $\delta \nu = 107.52$ kHz (channel-averaged spectral resolution of eight SKA-Mid channels). The analysis presented here only accounts for cosmic variance. The detectability of the local dimensions is studied in the presence of thermal noise in \cite{dosibhatla_2025_lss-morphology}. Filamentary features in intensity maps are captured by large $k_\parallel$ modes, which are less foreground-dominated than small $k_\parallel$ modes lying in the foreground wedge. Therefore, the detectability of the local dimensions is not expected to be significantly impacted by residuals of continuum foregrounds.

As gravitational clustering makes the late-time matter field highly non-Gaussian, and it also leads to the formation of filaments, sheets, nodes, and voids, identifying these building blocks can yield information on the non-Gaussianity in the matter field. The local dimension distribution can also distinguish between different astrophysical scenarios. Therefore, local dimension can be employed as a higher-order summary statistic to constrain astrophysics and also cosmology, as different cosmological models lead to different gravitational clustering patterns.


\section{Discussion}

In this chapter, we explored and demonstrated the detectability of various higher-order statistics of the 21-cm LIM, beyond the traditional two-point correlation function and power spectrum, using the future SKA-Mid. We observe that different higher-order statistics exhibit complementary sensitivities to cosmological parameters, physical processes, and systematic effects, as they can quantify the inherent non-Gaussianity in the signal, which is mostly driven by the non-linearity in the matter distribution. While some one-point statistics can effectively break degeneracies in cosmological parameters and constrain the HI luminosity function, they are susceptible to calibration uncertainties. On the other hand, the bispectrum can potentially provide shape-dependent constraints on HI-dark matter bias parameters and redshift-space distortions, whereas image-domain morphological statistics can, in principle, directly capture the topology of the cosmic web, although this is highly dependent on the resolution of the 21-cm maps. This effectively suggests that joint constraints by multiple higher-order statistics might enhance the reliability and robustness of these constraints. Such joint constraints might also help validate that the inferred parameters are not artifacts of particular systematic effects.

While higher-order statistics offer significantly enhanced constraining power, they also present substantial challenges: theoretical and numerical modeling becomes increasingly complex and often requires forward-modeling approaches or emulator-based pipelines calibrated on large-volume simulations. Covariance estimation is computationally expensive and often requires either a large ensemble of mock observations or analytical approximations with perturbative corrections. Moreover, systematics such as beam effects, foreground residuals, and survey geometry must be carefully accounted for to avoid bias in parameter inference exercises using 21-cm maps.

Despite these challenges, the information gained from including higher-order statistics in the inference pipeline is compelling. Once next-generation experiments, such as SKA-Mid, are operational, these techniques will become not just useful but necessary for achieving the full scientific potential of 21-cm LIM. This chapter consolidates current efforts within the SKA-LIM working group to advance these methodologies and provides a roadmap for future studies aiming to incorporate higher-order statistics into the LIM cosmology pipeline.


\section*{Author List Ordering}
Authors for this chapter are ordered according to their overall level of contribution, in line with that expected for a small author list publication.

\section*{Acknowledgment}
SM, MMD, LN, and AD acknowledge financial support through the Core Research Grant titled ``Observing the Cosmic Dawn in Multicolour using Next Generation Telescopes'' from the Science
and Engineering Research Board (SERB) and the Department of Science and Technology (DST), Government of India. SM, MV, AD, LN, SKP and MMD acknowledge financial support through the project titled “Illuminating the Dark Sector of the Cosmos in the SKA Era” (Project No. P3497) funded under the “Scheme for Promotion of Academic and Research Collaboration (SPARC)” from the Ministry of Education, India.  LN acknowledges support from the Abdus Salam International Centre for Theoretical Physics (ICTP) under the `ICTP Sandwich Training Educational Programme (STEP)’  SMR.3991 and SMR.4129. SKP, LN, and YM acknowledge the financial support by the Department of Science and Technology, Government of India, through the INSPIRE Fellowship.
DS acknowledges the support of the Canada 150 Chairs program, the Fonds de recherche du Québec Nature et Technologies (FRQNT) and the Natural Sciences and Engineering Research Council of Canada (NSERC) joint NOVA grant, and the Trottier Space Institute Postdoctoral Fellowship program. BVG and CU were supported by the European Union (ERC StG, LSS\textunderscore BeyondAverage, 101075919). JLB acknowledges funding from the grant UC-LIME (PID2022-140670NA-I00), financed by MCIN/AEI/ 10.13039/501100011033/FEDER, UE. 
AKS acknowledges the support from the National Science Foundation (grant No. 2206602). CH's work is funded by the Volkswagen Foundation, and supported through Germany's Excellence Strategy through EXC$\sim$2181/1 -- 390900948 (the Heidelberg STRUCTURES Excellence Cluster) and with support by the Federal Ministry of Education and Research (BMBF) and the Ministry of Science, Research and the Arts of Baden-Württemberg. PG’s work was supported by the TITAN ERA Chair project (contract no. 101086741) within the Horizon Europe Framework Program of the European Commission and the  Agence Nationale de la Recherche (ANR-22-CE31-0014-01 TOSCA).

\bibliographystyle{abbrvnat-maxbibnames4}
\bibliography{chapter} 



\end{document}